\documentclass[pra,twocolumn,superscriptaddress,longbibliography]{revtex4-2}
\usepackage{graphicx}
\usepackage{dcolumn}
\usepackage{bm}
\usepackage{physics}
\usepackage{amsmath}
\usepackage{array}
\usepackage{color}
\usepackage{times}
\newcolumntype{P}[1]{>{\centering\arraybackslash}p{#1}}
\usepackage[colorlinks=true, breaklinks=true, linkcolor=red, citecolor=blue, urlcolor=blue]{hyperref}

\newcommand{\bonnpi}{Physikalisches Institut, University of Bonn, Nussallee 12, 53115 Bonn, Germany}
\newcommand{\geneva}{Department of Quantum Matter Physics, University of Geneva, Quai Ernest-Ansermet 24, 1211 Geneva, Switzerland}

\newcommand{\psich}{Laboratory for Theoretical and Computational Physics, Paul Scherrer Institute, CH-5232 Villigen, Switzerland}
\newcommand{\epfl}{Institute of Physics, Ecole Polytechnique F\'ed\'erale de Lausanne (EPFL), CH-1015 Lausanne, Switzerland}
\usepackage{tikz}
\usetikzlibrary{calc}
\usetikzlibrary{shapes}
\usepackage{ifthen}
\usepackage{booktabs}
\usepackage{multirow}

\begin{document}

\title{Level statistics of the one-dimensional ionic Hubbard model}
\date{\today}

\begin{abstract}
  In this work we analyze the spectral level statistics of the one-dimensional ionic Hubbard model, the Hubbard model with an alternating on-site potential. In particular, we focus on the statistics of the gap ratios between consecutive energy levels. This quantity is often used in order to signal whether a many-body system is integrable or chaotic. A chaotic system has typically the statistics of a Gaussian ensemble of random matrices while the spectral properties of the integrable system follow a Poisson statistics. We find that whereas the Hubbard model without alternating potential is known to be integrable and its spectral properties follow a Poissonian statistics, the presence of an alternating potential causes a drastic change in the spectral properties which resemble the one of a Gaussian ensemble of random matrices.  However, to uncover this behavior one has to separately consider the blocks of all symmetries of the ionic Hubbard model. 
\end{abstract}
\author{Jeannette De Marco}
\affiliation{\bonnpi}
\author{Luisa Tolle}
\affiliation{\bonnpi}
\author{Catalin-Mihai Halati}
\affiliation{\geneva}
\author{Ameneh Sheikhan}
\affiliation{\bonnpi}
\author{Andreas M. L\"auchli}
\affiliation{\psich}
\affiliation{\epfl}
\author{Corinna Kollath}
\affiliation{\bonnpi}

\maketitle

\section{Introduction \label{sec:intro}}

In the past years a significant effort has been devoted to the understanding of the dynamics of quantum systems. In contrast to the equilibrium physics at low temperature which is typically dominated by the low energy properties of a quantum system, the non-equilibrium physics relies often on properties of an arbitrary part of the spectrum.  A central question is the determination of the conditions under which an isolated quantum many-body system thermalizes or fails to thermalize \cite{Srednicki1994,KollathAltman2007, PolkovnikovVengalattore2011,NandkishoreHuse2015,AbaninSerbyn2019}. In this context it is generally believed that a generic chaotic system is best suited to exhibit thermalization towards a suitable statistical mechanics ensemble~\cite{Haake2000}.

Recent studies have put forward several examples of non-equilibrium phenomena, which provide examples of systems failing to thermalize, such as many-body localization \cite{BaskoAltshuler2006,GornyiPolyakov2005,PalHuse2010,NandkishoreHuse2015,AbaninSerbyn2019}, and Hilbert space fragmentation~\cite{SalaPollmann2020, KhemaniNandkishore2020}. These systems feature a large number of (almost) conserved quantities and are thus not chaotic.
The phenomenon of quantum many-body scarring \cite{ShiraishiMori2017,TurnerPapic2018,KhemaniChandran2019,HoLukin2019} highlights that even in overall chaotic systems
tiny subspaces can be found, which fail to thermalize. This scenario is especially relevant if the initial conditions lie in this subspace~\cite{BernienLukin2017}.

A powerful probe of the properties of a many-body quantum system in the context of quantum chaos are the universal properties of the level statistics. There exist two cornerstone limits: the Poisson statistics and the statistics stemming from random matrix theory \cite{Mehtabook}.
The random matrix theory statistics is conjectured to hold for generic chaotic systems \cite{BerryZiman1977, BohigasSchmit1984}, where the energy levels repel each other. The exact class of random matrices is determined by the symmetries present in the system.  In contrast, the Poisson statistics is found for  quantum integrable systems. 

The level statistics of different many-body quantum systems have been classified as for example of the prime examples of the Hubbard model~\cite{PoilblancMontambaux1993} and variants of it~\cite{JafariHosseinzadeh2020}, or its bosonic counterpart~\cite{KollathLaeuchli2010}, a kicked-parameter model of spinless fermions~\cite{Prosen1999},
or more recently a family of Sachdev-Ye-Kitaev models~\cite{Haque2019}.

The Hubbard model is one of the most studied models in condensed matter physics, since it is one of the simplest models containing the competition between the kinetic term and the interaction term of fermions. A large amount of work has been devoted to study the properties of the Hubbard model \cite{Gebhard1997} and in particular of its one-dimensional version~\cite{EsslerKorepin2005, Giamarchibook}. The one-dimensional model has the particularity that it is Bethe-Ansatz integrable which enabled a lot of detailed studies~\cite{EsslerKorepin2005}. In agreement with the conjecture mentioned before, the level statistics of the one-dimensional Hubbard model was found to follow the Poisson statistics \cite{PoilblancMontambaux1993}, while e.g.~the two-dimensional Hubbard model at low filling exhibits the statistics of Gaussian orthogonal ensemble (GOE)~\cite{Bruus1996}.

In this work, we will investigate the spectral properties of a generalization of the one-dimensional Hubbard model by adding an alternating local potential term. 
Such an ionic Hubbard model has been realized in cold atomic gases using a superlattice potential \cite{PertotKollath2014,MesserEsslinger2015} and proposed to describe GeSe \cite{KennesRubio2020}. Further, the ionic Hubbard model has been devised in order to study the physics occurring at neutral-ionic phase transitions as they occur in solids for example, ionic to neutral transitions in organic charge-transfer solids \cite{TorranceLaPlaca1981, NagaosaTakimoto1986} and at ferroelectric transitions in perovskites \cite{EgamiTachiki1993}. An alternating local potential is also at the heart of the staggered fermion formulation of massive fermions in quantum field theories on a lattice~\cite{Kogut1975}, and is of relevance to ongoing efforts to simulate quantum field theories.

 Such an alternating  potential is widely believed to break the integrability of the Hubbard model. However, recently, numerical results pointing towards a Poisson statistics of the levels \cite{JafariHosseinzadeh2020} were reported. The tension with the general expectation motivated us to investigate the level statistics properties of this model in more detail. Indeed we find that after all the symmetries of the Hamiltonian are taken into account, the level statistics are clearly of GOE nature in a wide range of parameters considered. 
  
In Sec.~\ref{sec:model} we describe the one-dimensional ionic Hubbard model considered in this work. In Sec.~\ref{sec:statistics} we present the level statistics approach we use to show the chaotic character of the ionic Hubbard model. In order to employ this procedure we first need to identify all symmetries of the model, which we present in Sec.~\ref{sec:sym}. We determine the eigenenergies of the ionic Hubbard model by performing complete numerical exact diagonalization using the identified symmetries. Our results regarding the spectral properties for generic and half filling are presented in Sec.~\ref{sec:results}.

\section{Model \label{sec:model}}

The Hubbard model is one of the standard models of condensed matter physics. It is the simplest model which describes the competition between the  motion of two species of fermions, called spin up and spin down fermions, in a periodic lattice potential and their on-site interaction. The one-dimensional Hubbard model is given by the Hamiltonian 
\begin{align}
 H_0 = &-J \sum_{j=1}^{L} \sum_\sigma (c_{j,\sigma}^\dagger c_{j+1,\sigma} + c_{j+1,\sigma}^\dagger c_{j,\sigma})\nonumber\\
    &+ U \sum_{j=1}^L n_{j,\uparrow}n_{j,\downarrow}
\end{align}
Here $c_{j,\sigma}$ are the fermionic operators describing a fermion with spin $\sigma =\uparrow , \downarrow$ on site $j$. $J$ is the tunneling amplitude of the fermions from one lattice site to the neighboring ones, and $U$ the strength of the on-site interaction between the different fermionic species. $L$ denotes the length of the chain, which is chosen to be an even number. The total particle number of the fermions is denoted by $N$. We note that we focus on even $N$, therefore the total spin is integer. We use in the following periodic boundary conditions, i.e.~we identify $c^{(\dagger)}_{L+1,\sigma}\equiv c^{(\dagger)}_{1,\sigma}$.
Depending on the parameters chosen, the one-dimensional Hubbard model has many interesting phases~\cite{Giamarchibook,EsslerKorepin2005}. These reach from a metallic/Luttinger liquid phase ($U=0$, and $U>0$ away of half filling) over a band insulating phase for the totally filled case, a Mott insulating phase at half filling for $U>0$ to a superfluid phase for the attractive interaction $U<0$. Here we mentioned only the phases occuring for a spin-balanced situation. 

One experimentally relevant extension of the Hubbard model is the so-called ionic Hubbard model which has an additional alternating local potential. Its Hamiltonian is given by
  \begin{align}
  \label{eq:ham_ionic}
 H = &H_0 -\eta \sum_{j=1}^L \sum_\sigma (-1)^j n_{j,\sigma},
  \end{align}
where $\eta$ gives the strength of the alternating potential.
The ground state phase diagram of the ionic Hubbard model has attracted a lot of interest \cite{FabrizioNersesyan1999, FabrizioNersesyan2000, GidopoulosTosatti2000, TorioCeccatto2001,WilkensMartin2001, KampfBrune2003, ZhangLin2003, ManmanaSchonhammer2004, BatistaAligia2004, OtsukaNakamura2005, RefolioRubio2005, AligiaBatista2005, LegezaSolyom2006, TincaniBaeriswyl2009, HafezJafari2010, GoJeon2011, KimJeon2014, HafezTorbatiUhrig2014, HafezTorbatiUhrig2015, BagKrishnamurthy2015, LoidaKollath2017, ChattopadhyayGarg2019} and many interesting phases have been pointed out. In particular, the very interesting phase of a bond order arises. 
These results have been obtained using either approximative methods or numerical methods because of the common believe that the ionic Hubbard model 
for generic couplings (i.e.~away from the known integrable or free parameter sets) is not Bethe-Ansatz solvable.

\section{Level Statistics \label{sec:statistics}}

In this section we would like to summarize some properties of the spectra of quantum many-body models. It has been conjectured that the spectral statistics of quantum many-body systems displays either  universal features described by random matrix theory (RMT) for chaotic quantum systems \cite{BerryZiman1977, BohigasSchmit1984, BohigasGiannoni1986, GuhrWeidenmueller1998} (if all known symmetries are removed), or follows Poisson statistics for quantum integrable systems.
There exist numerous examples in the literature that show the usefulness of the spectral analysis in order to obtain information on the chaoticity or integrability of many-body quantum models \cite{PoilblancMontambaux1993, HsuAngle1993, MontambauxSire1993, OganesyanHuse2007, KollathLaeuchli2010, PalHuse2010, SerbynMoore2016, Prosen1999, Haque2019}.

In order to quantify the spectral properties of a model one important quantity is the distance between adjacent many-body eigenvalues
  \begin{align}
 \delta_n=E_{n+1}-E_n,
  \end{align}
where $\{E_n\}$ are the eigenvalues of the Hamiltonian sorted in an ascending order. In the case of integrable models, for which one can find an extensive number of conserved quantities, the distribution of the level spacings should follow a Poisson distribution
\begin{align}
P(\delta/\Delta)=\exp\left(-\frac{\delta}{\Delta}\right),
\end{align}
where $\Delta$ is the mean level spacing. In contrast, for a chaotic model the underlying symmetries, as the time-reversal symmetry, determine with which random matrix ensemble the model shares the same universal features. Similar to Hubbard model the ionic Hubbard Hamiltonian considered here (Eq.~\ref{eq:ham_ionic}) is invariant under time reversal operator $T$ and has rotational symmetry which leads to the Gaussian orthogonal ensemble (GOE) in random matrix theory~\cite{GuhrWeidenmueller1998}.

For the GOE the level spacing distribution has the Wigner-Dyson form
\begin{align}
P(\delta/\Delta)=\frac{\pi}{2}\frac{\delta}{\Delta}\exp\left(-\frac{\pi}{4}\frac{\delta^2}{\Delta^2}\right).
\end{align}

Numerically, often an alternative way is employed in order to characterize the spectral properties. One considers the behavior of the gap ratio for consecutive levels defined by \cite{OganesyanHuse2007}
\begin{align}
r_n=\frac{\min\left(\delta_n,\delta_{n+1}\right)}{\max\left(\delta_n,\delta_{n+1}\right)}.
\end{align}
This approach has the advantage of bypassing the need to compute the density of states. The computation of the density of states would imply a procedure for the unfolding of the spectrum, which can lead to inaccurate numerical results \cite{GomezRetamosa2002}.
The probability distribution of the consecutive gap ratios for the Poisson statistics is given by 
\begin{align}
\label{eq:Poisson_dist}
P_\text{Poisson}(r)=\frac{2}{(1+r)^2},
\end{align}
and for the GOE statistics by \cite{AtasRoux2013}
\begin{align}
\label{eq:GOE_dist}
P_\text{GOE}(r)=\frac{27}{4}\frac{r+r^2}{\left(1+r+r^2\right)^{5/2}}.
\end{align}
One can quantify the proximity of the computed distribution of gap ratios to Poisson or GOE by using the mean value $\langle r \rangle$ or the value $P(r=0)$. The expected values for the Poisson and GOE distributions are given in Table~\ref{tab:values}.

\begin{table}[ht!]
  \centering
  \begin{tabular}{|P{2.5cm}|P{2cm}|P{2cm}|}
    \hline
  Distribution & $\langle r \rangle$ & $P(r=0)$ \\ 
  \hline
  \hline
  GOE ($m=1$) & $0.536$ & $0$ \\ 
  \hline
  GOE ($m=2$) & $0.423$ & $1.408$ \\ 
  \hline
  GOE ($m=3$) & $0.403$ & $1.715$ \\ 
  \hline
  Poisson & $0.386$ & $2$ \\ 
  \hline
\end{tabular}
  \newline\newline
  \caption{Values of averages $\langle r \rangle$  and the probability at $r=0$, for the GOE distribution for different number of symmetry blocks $m$ (of equal size) and the Poisson distribution. Values taken from  Refs.~\cite{AtasRoux2013, GiraudAlet2020}.}\label{tab:values}
\end{table}

The comparison of the spectral properties of a quantum chaotic Hamiltonian with the ones of the RMT is typically performed within each symmetry sector of the symmetries present in the Hamiltonian. These symmetries comprise for example spatial translations and reflections, but also more special symmetries as particle hole transformations. In the case in which a few symmetries are known for the Hamiltonian, one typically block diagonalizes the Hamiltonian in order to perform the analysis of the level statistics. However, in certain situations the symmetries might not be known or the block diagonalization procedure may be impractical or not possible. Several work investigated different situations where additional symmetries are present \cite{RosenzweigPorter1960,BerryRobnik1984,SunLiu2020,GiraudAlet2020}.

In particular, Giraud and coworkers \cite{GiraudAlet2020} present analytical results for the distributions of random matrices comprised of several independent symmetry blocks, corresponding to the presence of a discrete symmetry.
In particular, we make use of the results of Ref.~\cite{GiraudAlet2020} for the GOE distribution for $m=2~\text{or}~3$ blocks of equal size, to show the chaotic nature of the half-filled ionic Hubbard model.
The mean value $\langle r \rangle$ and the value $P(r=0)$ for the GOE distributions with $m=2~\text{or}~3$ are given in Table~\ref{tab:values} and the distributions are plotted in Fig.~\ref{fig:halffilling}. The analytical expression of the distributions can be found in the Supplementary Material of Ref.~\cite{GiraudAlet2020}. 
One can observe that by having multiple symmetry sectors the value $P(r=0)$ becomes finite and increases for larger $m$ (see Table~\ref{tab:values}). The mean value $\langle r \rangle$ is already closer to the value expected for the Poisson distribution, Eq.~(\ref{eq:Poisson_dist}), than the one for the GOE distribution, Eq.~(\ref{eq:GOE_dist}), even for just $m=2~\text{or}~3$ blocks. 
Whereas the distribution of consecutive level spacings for $m=2$ shows very pronounced leveling off at low values as $r\to 0$, the distribution of $m=3$ already shows a behavior very similar to the Poisson distribution. By taking a very large number of symmetry sectors, $m\to\infty$, one will recover the Poisson distribution \cite{GiraudAlet2020}. Thus, extreme care has to be taken in the numerical analysis of the spectral properties of many body quantum model, since missing just a few symmetries the numerical results for a chaotic system could resemble an integrable one. We demonstrate this explicitly for the ionic Hubbard model in the Appendix~\ref{app:withoutS}.

We note that in the following we mainly make use of the results for the GOE distributions for which the $m$ symmetry blocks are of equal size. However, one can  perform the same analysis also for the case of unequal blocks \cite{GiraudAlet2020}, as we use it in Appendix~\ref{app:withoutS}. In the case of $m=2$, as the ratio of the sizes of the two blocks goes from $0$ to $1$ the values $\langle r \rangle$ and $P(r=0)$ interpolate between the values for $m=1$ and $m=2$.

\section{Symmetries \label{sec:sym}}

In order to study the level statistics of the ionic Hubbard model we need to identify the discrete symmetries existing in the model. In the following we list the symmetries (beside the time-reversal symmetry discussed already in the previous section) occurring in the ionic Hubbard model with periodic boundary conditions and comment on how we take them into account.

\paragraph{Gauge symmetry:} The global $U(1)$ gauge symmetry
\begin{equation}
U c_{j,\sigma} U^\dagger= e^{i\phi}c_{j,\sigma}, 
\end{equation}
where $\phi$ is a real number, leads to the conservation of the particle number $N_\sigma$ with spin $\sigma$ and consequently also of the total particle number $N=N_\uparrow+N_\downarrow$. This conservation is directly implemented in the numerical exact diagonalization method. 

\paragraph{Spin rotation symmetry:}
 The $SU(2)$ spin rotation symmetry, well known from the Hubbard model, also exists in the presence of an alternating potential. The spin operators are defined by
\begin{equation}
S^\alpha =\frac{1}{2}\sum_{j=1}^L\sum_{a,b\in{\uparrow,\downarrow}}c_{j,a}^\dagger (\sigma^\alpha)_{a,b}c_{j,b},\quad \alpha=x,y,z
\end{equation}
where $\sigma^\alpha$ are the Pauli matrices. These spin operators represent a $\text{SU}(2)$ Lie algebra and generate the rotations in spin space. As all spin operators commute with the Hamiltonian of the ionic Hubbard model in Eq.~(\ref{eq:ham_ionic}), $[H,S^\alpha]=0$. Thus, the Hamiltonian is rotationally invariant in spin space which corresponds to the conservation of the total spin $S$. The symmetry sectors corresponding to this symmetry are labeled by $\ket{s, m_z}$, where $S^2\ket{s,m_z}=s(s+1)\ket{s,m_z}$ and $S^z\ket{s,m_z}= m_z \ket{s,m_z}$ (where $\hbar=1$). The conservation of the $S^z$ component is directly implemented in the numerical exact diagonalization method. However, the symmetry sectors corresponding to the absolute spin $S$ are reconstructed from the numerical results. For more explanation on how this is performed see Appendix~\ref{app:impl_symm}.
\paragraph{Translational symmetry:} 
The lattice in the ionic Hubbard model has a two site unit cell. Therefore, a translational symmetry with respect to a translation by an even number of lattice sites exists, i.e. 
\begin{equation}
T_x c_{j,\sigma}T_x ^\dagger= c_{j+2,\sigma}.  
\end{equation}
The corresponding conserved (dimensionless) momentum $k$ determine the eigenvalues $\exp(ik)$ of the generator $T_x$. The possible momentum values are $k=\frac{2\pi}{L/2} j$ with $j=0,1,\dots,\frac{L}{2}-1$, for a chain with an even number $L$ of sites. Thus, each symmetry sector is characterized by its dimensionless momentum $k$. 
\paragraph{Parity (reflection) symmetry:}
In the ionic Hubbard model there exists a reflection symmetry given by the transformation 
\begin{equation} 
 P c_{j,\sigma}P^\dagger= c_{2-j,\sigma}. 
 \end{equation}
This transformation also leaves the Hamiltonian invariant. Here, the reflection center is chosen as the first site of the chain. Due to the translational symmetry,  the reflection center can be chosen on any site of the chain (but not on a bond center). The eigenvalues corresponding to the reflection symmetry are $p=\pm1$.

Note that all symmetry generators besides $(T_x,P)$ commute.
As the reflection symmetry changes the direction of the momentum, it does not commute with the translational symmetry in general, but only if the momentum is $k=0$ or $k=\pi$. The momentum $k=\pi$ only exists if $\frac{L}{4}$ is an integer number. Therefore, we cannot diagonalize the matrix with respect to both the translation and reflection symmetry beside in the special case of $k=0$ or $k=\pi$.
We choose to use the translation symmetry first and the reflection symmetry only for the special case of $k=0,\pi$. 
 
Combining all symmetries, the individual symmetry blocks are thus characterized by $(N, s, m_z, k, p)$. The last quantum number is only present for $k=0,\pi$. 
Since the level statistics are independent of the respective symmetry sectors, if not stated otherwise, the distributions are calculated within all computable sectors individually and are only afterwards combined to decrease the statistical error of the distributions.

\paragraph{Particle-hole like symmetry:}
In the ionic Hubbard model an additional symmetry is present at half-filling. 
This additional symmetry resembles a particle hole symmetry, here between two sublattices, i.e. with the transformation  
\begin{equation}
  \label{eq:particle_hole}
C c_{j,\sigma} C^\dagger = (-1)^j c^\dagger_{j+1,\sigma}.
\end{equation}
 The ionic Hubbard Hamiltonian at half filling is invariant under this transformation. 
 We discuss in the result section how this symmetry influences the properties of the level statistics.

\paragraph{Symmetries used in exact diagonalization}

In the exact diagonalization code we build subsectors for fixed $N$, $m_z$, $k$ and reflection quantum number, if applicable. It is tedious to combine the particle hole-like symmetry with the other symmetries in the existing code, and we did therefore not implement it. However, in the subsequent analysis we are able to detect the presence of
this symmetry in the resulting level statistics distributions and to account for it in the analysis, and additionally we found a way to explicitly break the particle hole-like symmetry by alternating Hubbard on-site interactions, while retaining all other symmetries. 

In the following we diagonalize individual blocks of sizes up to $ \sim 1.27\times 10^4$. For the averages over all the symmetry blocks up to $ \sim 3.84\times 10^5$ eigenvalues were taken into account.

\section{Results}
\label{sec:results}

In this section we show our results on the spectral properties of the ionic Hubbard model and an extension using alternating interactions. We start discussing the results at quarter filling ($N/L=1/2$) and at filling $N/L=2/3$ as typical fillings (Sec.~\ref{sec:away_half_filling}), only the standard symmetries (Sec.~\ref{sec:sym} (a)-(d)) are present for this case. We recover the GOE like behavior which points towards the chaotic character of the ionic Hubbard model. 
Additionally, we discuss in Sec.~\ref{sec:half_filling} the case of half filling, which is special due to the extra particle hole like symmetry (Sec.~\ref{sec:sym} (e)). We show that also in this case, the results are in agreement with a GOE like behavior occurring within the different symmetry sectors. 

\subsection{Away from half filling}
\label{sec:away_half_filling}
We start considering the ionic Hubbard model, Eq.~(\ref{eq:ham_ionic}), at quarter filling, i.e. $N/L=1/2$. We compute the energy levels of the Hamiltonian in each block of the symmetries discussed in Sec.~\ref{sec:sym} and consider the level distribution separately within these blocks. The data obtained from different symmetry sectors at quarter filling is then assembled in a single histogram. In the representation of the histograms the chosen size of the bins plays a role. We choose the bin size in order to minimize the statistical fluctuations and still obtain a good representation of the distribution. Since the distributions vary depending on the Hamiltonian parameters, we adapted the bin size correspondingly.

\begin{figure}[h!]
\centering
\includegraphics[width=0.95\linewidth]{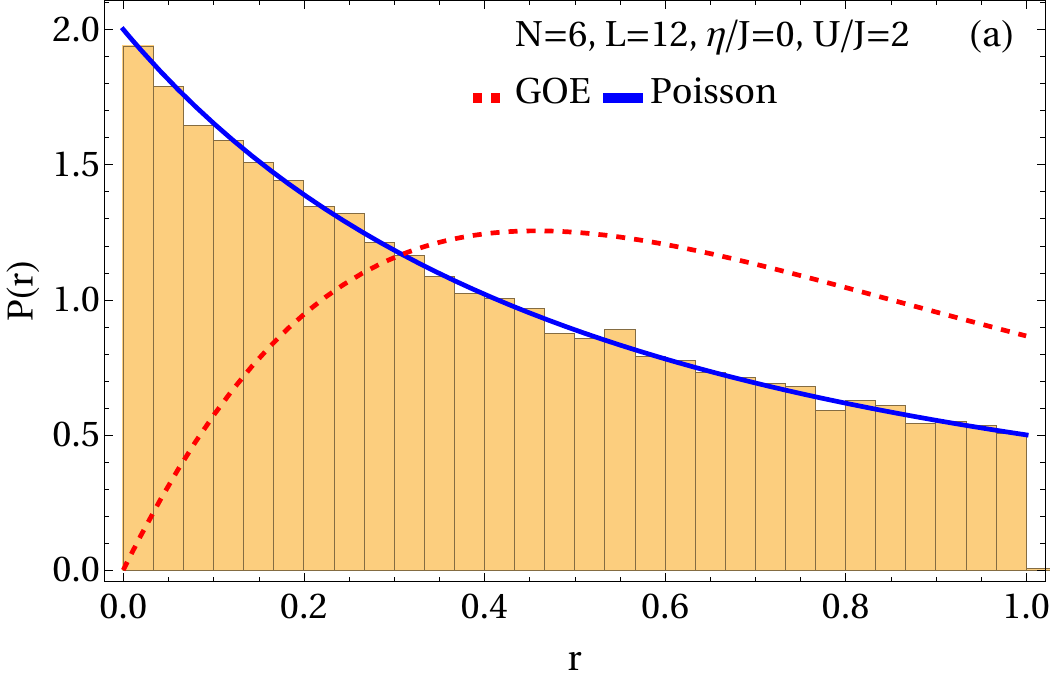}
\vspace{5mm}
\includegraphics[width=0.95\linewidth]{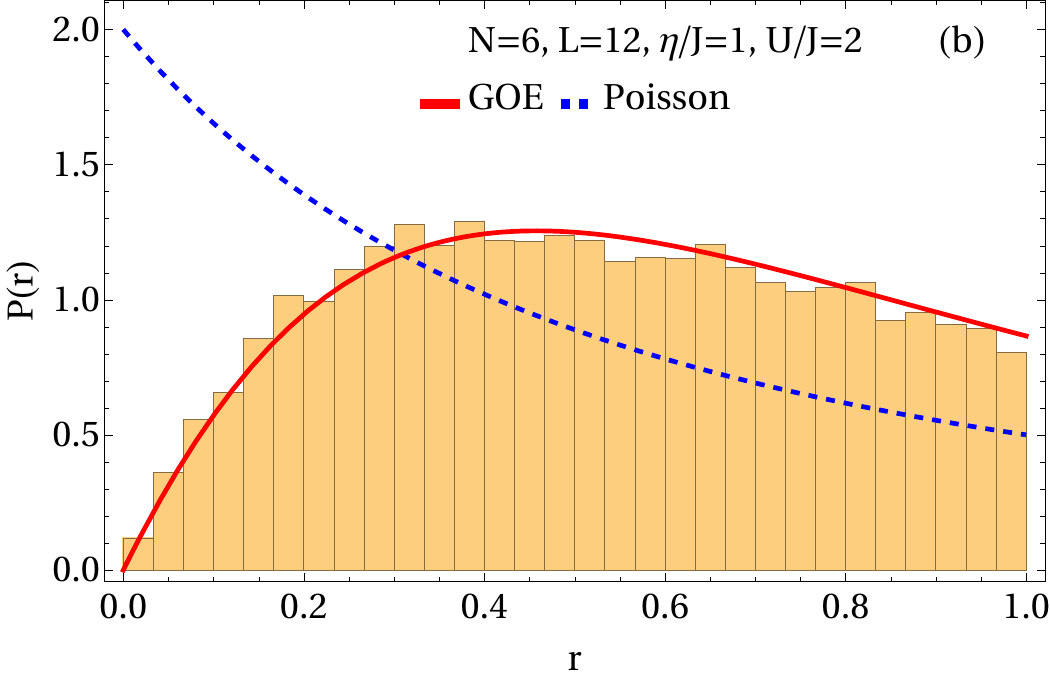}
\caption{
Distribution of the ratios of consecutive level-spacings for (a) the standard Hubbard model ($\eta/J = 0$) and (b) the ionic Hubbard model at $\eta/J = 1$ 
in a chain with $N = 6$, $L = 12$ at $U/J=2$ combining all $s$ and $k,p$ symmetry sectors for $m_z = 0$. We use $30$ bins to represent the distribution. 
}
\label{fig:distributions_eta}
\end{figure}

\begin{figure}[h!]
\centering
\includegraphics[width=0.95\linewidth]{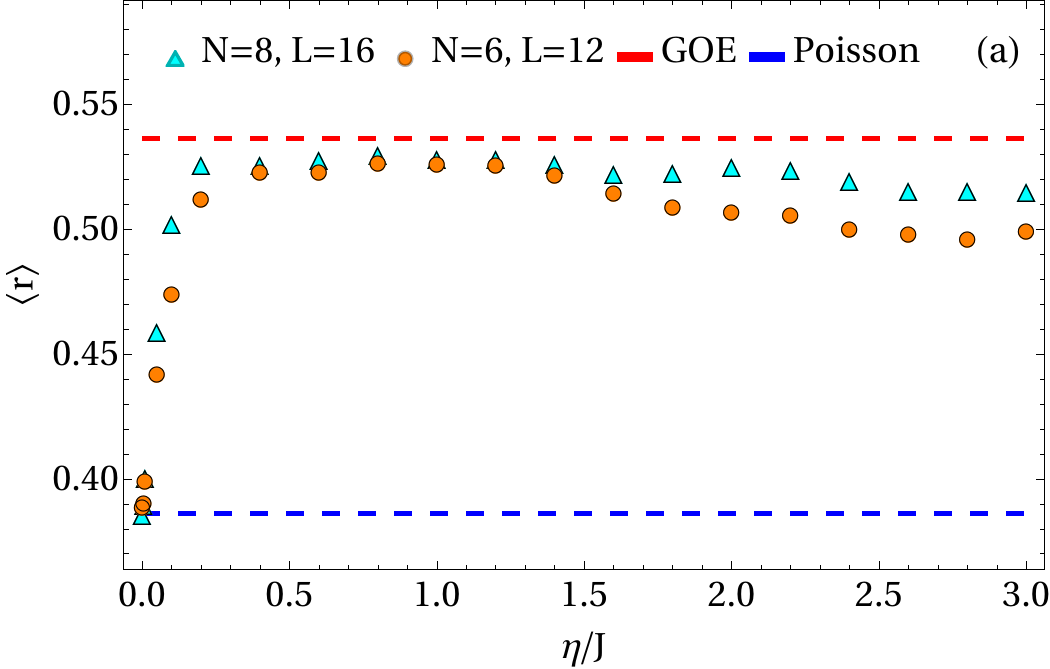}
\vspace{5mm}
\includegraphics[width=0.95\linewidth]{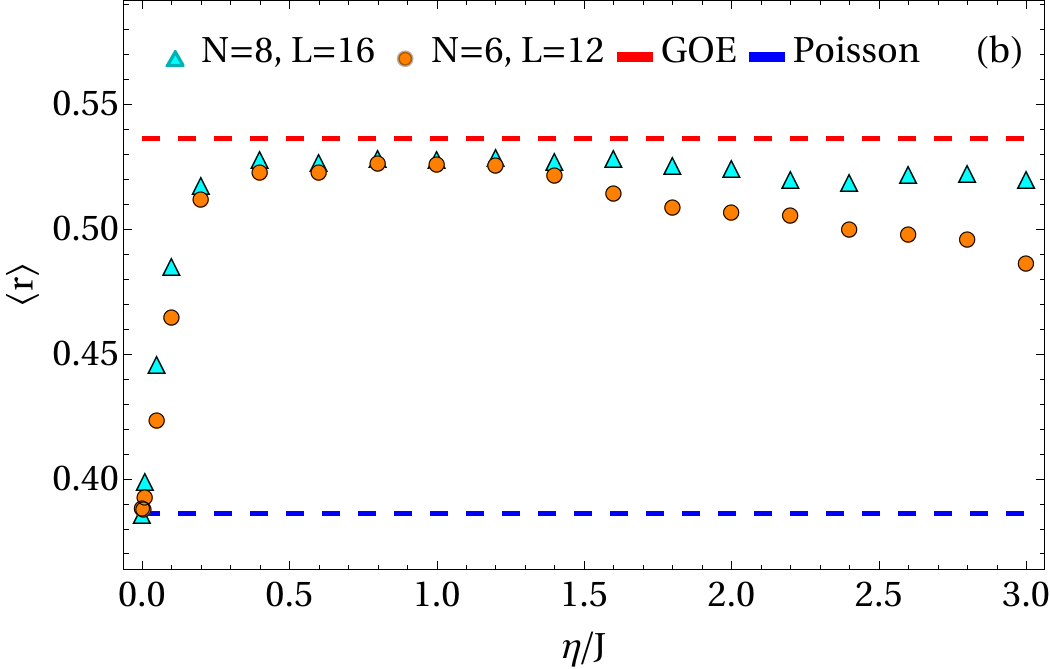}
\caption{Evolution of the mean value $\langle r \rangle$ of the ratios of consecutive level-spacings with the strength of the alternating potential $\eta/J$ for a) $U/J=2$ and b) $U/J=4$. The expected mean value for the Poissonian distribution and the GOE are marked with the horizontal dashed lines. 
For system size $L=12$ combining all $s$ and $k,p$ symmetry sectors, for $L=16$ combining $s=3,4$ and all $k,p$ for $m_z = 0$.}
\label{fig:r_varying_eta}
\end{figure}

\begin{figure}[h!]
	\centering
	\includegraphics[width=0.99\linewidth]{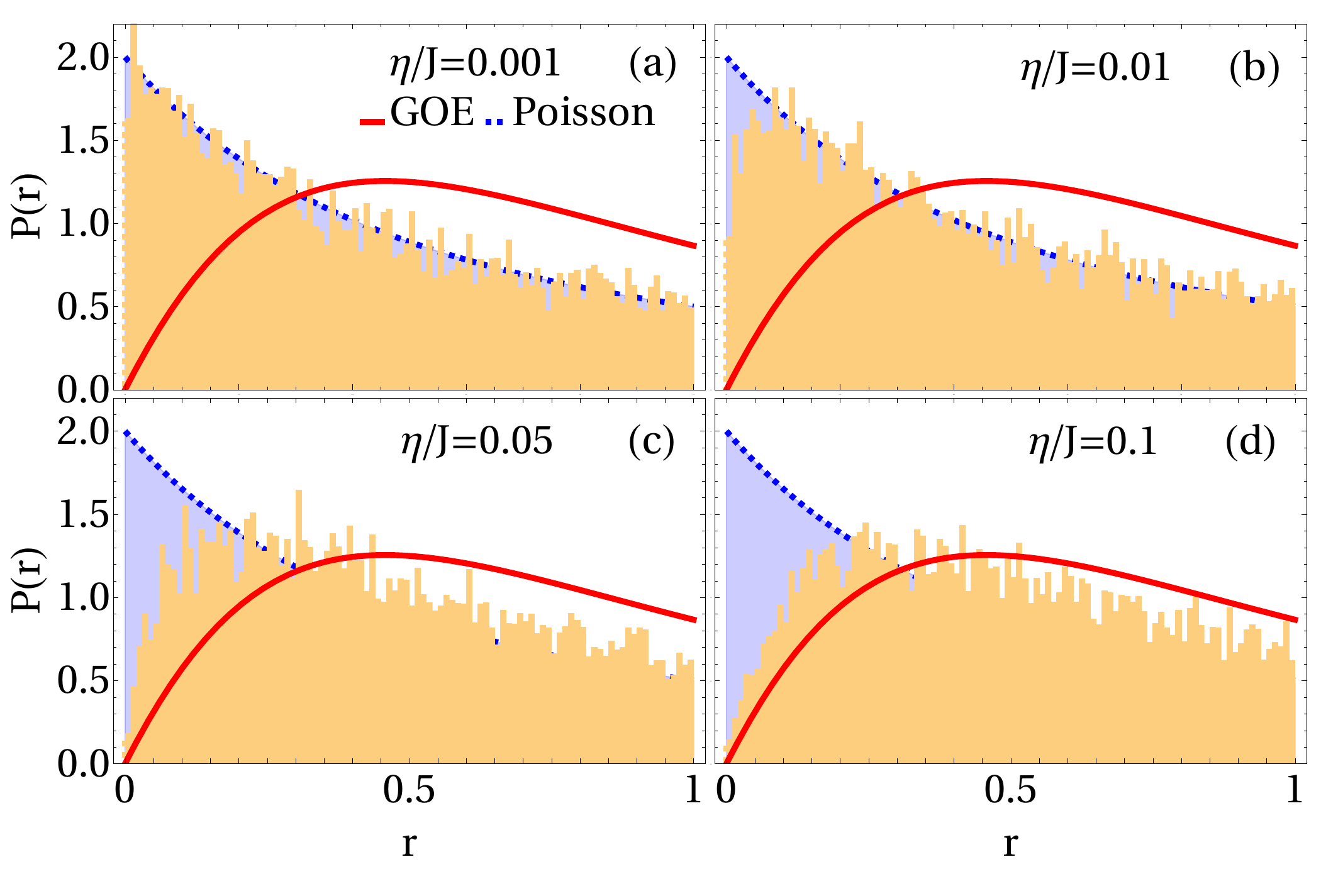}
	\caption{Distribution of the ratios of consecutive level-spacings for $N = 6$, $L = 12$ for the symmetry sector $m_z = 0$ combining  $s=0,1$ and all $k,p$ symmetry sectors. The parameters of the system are $U/J = 2$ and $\eta/J = 0.001, 0.01, 0.05, 0.1$ and $100$ bins. The mean values are given by $\langle r \rangle = 0.391, 0.405, 0.450, 0.482$. 
}
	\label{fig:distributions_small_eta}
\end{figure}

In Fig.~\ref{fig:distributions_eta}(a) we present the distribution of the ratios of consecutive level-spacings averaged over all symmetry blocks for the standard Hubbard model, i.e. $\eta/J=0$, as a benchmark of our procedure. For the standard Hubbard model, the ratios of consecutive level spacings are known to follow Poissonian level statistics \cite{PoilblancMontambaux1993}. For the chosen intermediate value of the interaction, $U/J=2$, we find a distribution which follows nicely the Poisson prediction, as can be
seen from the good overall agreement between the numerical histogram and the Poisson prediction for $P(r)$, including the correct limiting behavior as $r\rightarrow 0^+$. Furthermore the mean value $\langle r \rangle = 0.386$ of the numerical data, which does not depend on the number of bins, agrees with the Poisson distribution value in the first three digits.

Once we switch on the alternating potential of the ionic Hubbard model, the behavior is changed drastically. In Fig.~\ref{fig:distributions_eta}(b) the distribution $P(r)$ from the numerical data is shown for a value of the alternating potential of $\eta/J=1$ following closely the distribution expected for the GOE random matrix theory ensemble.  
The behavior for small $r$, where $P(r)$ clearly starts at 0 and rises linearly for small values of $r$, is in stark contrast with the Poisson behavior and is a consequence of
the level repulsion of chaotic systems. The mean value $\langle r \rangle= 0.515$ is also rather close to the GOE prediction $0.536$.

\begin{figure}[h!]
	\centering
	\includegraphics[width=0.99\linewidth]{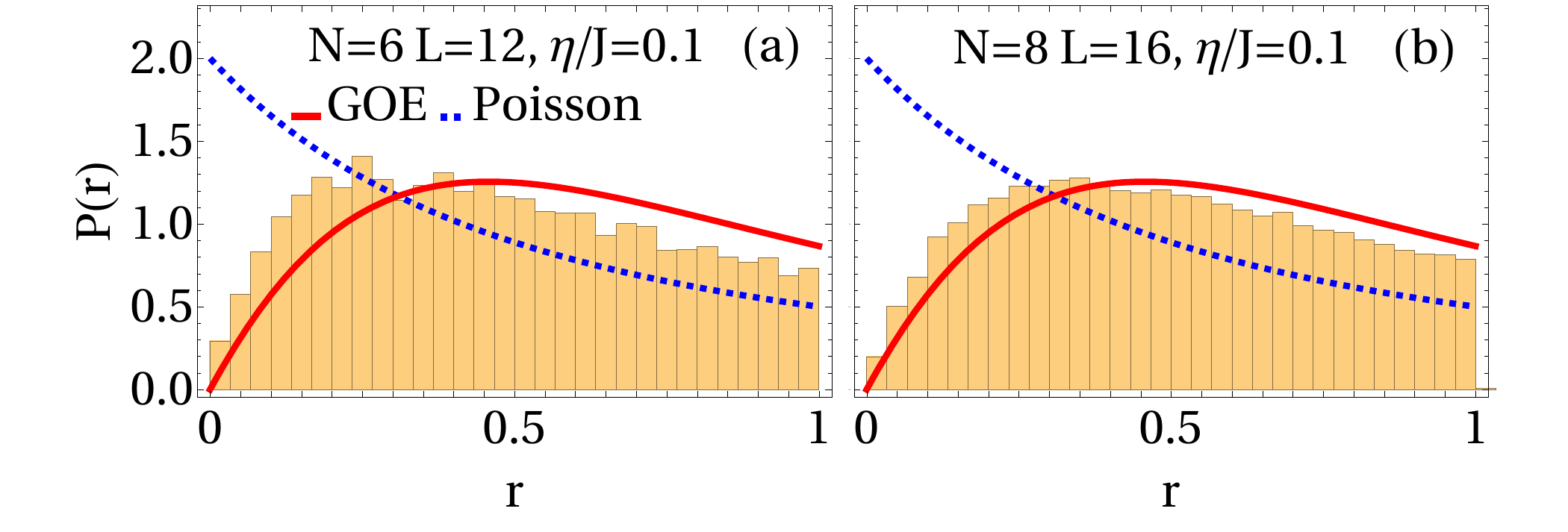}
	\caption{Distribution of the ratios of consecutive level-spacings for $N = 6$, $L = 12$,  $N = 8$, $L = 16$.
          The parameters of the system are $U/J = 1$ and $\eta/J = 0.1$, 30 bins. The mean value is given by $\langle r \rangle = 0.474, 0.502$. 
          For system size $N=6$ combining all $s$ and $k,p$ symmetry sectors, for $N=8$ combining $s=3,4$ and all $k,p$ for $m_z = 0$.
        }
	\label{fig:different_sizes}
\end{figure}

\begin{figure}[h!]
\centering
\includegraphics[width=0.95\linewidth]{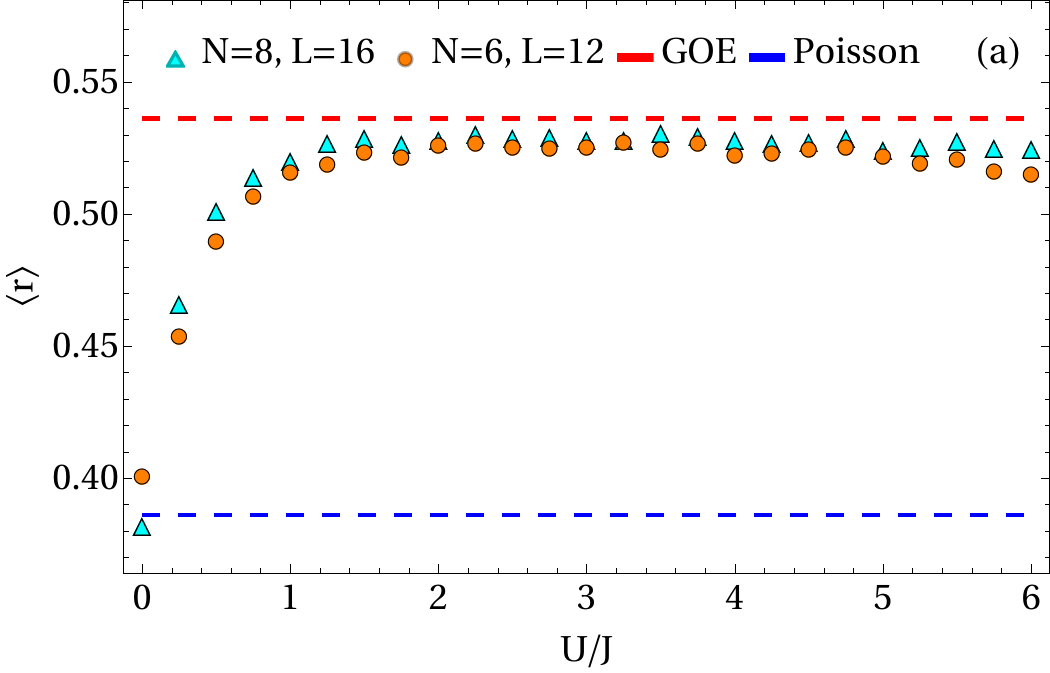}
\includegraphics[width=0.95\linewidth]{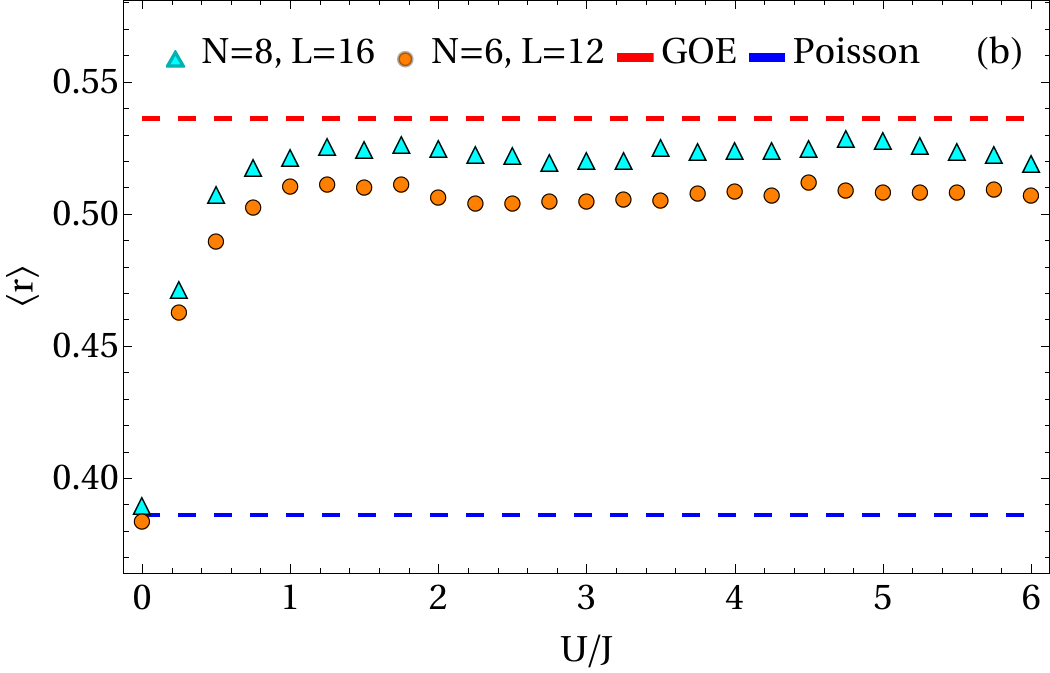}
\caption{Evolution of the mean value $\langle r \rangle$ of the ratios of consecutive level-spacings with the interaction strength $U/J$ for a) $\eta/J=1$ and b) $\eta/J=2$. The value of the Poisson distribution and the GOE are marked by horizontal dashed lines. 
  For system size $N=6$ combining all $s$ and $k,p$ symmetry sectors for $m_z=0$ and for $N=8$ combining $s=3,4$ and all $k,p$ for $m_z = 0$. The statistical standard error of the weighted mean of $r$ over all computed symmetry sectors was found to be of order $3\times10^{-4}$ or lower for all shown system sizes. This is below the symbol sizes.}
\label{fig:r_varying_U}
\end{figure}

In order to investigate in more detail the behavior as a function of the amplitude of the alternating potential $\eta/J$ and the dependence on $U/J$, we use the average value of the ratio of the consecutive level-spacings $\langle r \rangle$. 
For the Poissonian distribution the mean value of this ratio is given by $\langle r \rangle_{\text{P}} = 0.386$ and for the GOE the mean value is $\langle r \rangle_{\text{GOE}} = 0.536$ (see Table~\ref{tab:values}). The dependence on $\eta/J$ for fixed $U/J$ is shown in Fig.~\ref{fig:r_varying_eta} (a-c) for different interaction strength $U/J=2,4$. 
As discussed above, at $\eta=0$ the value of $\langle r \rangle$ is very close to $\langle r \rangle_{\text{P}}$ of the Poisson distribution. 
We observe that for the smallest non-zero values of $\eta$ we consider and for all considered interactions, the average value of $r$ increases very rapidly to almost the value expected for a GOE ensemble. A very good agreement with the expected value of the GOE is found in particular for $0.5\lesssim\eta/J\lesssim2$.

Since, in principle one expects an infinitesimal value of $\eta$ in the thermodynamic limit to follow the GOE like distributions, we investigate the transition from the Poisson distribution at $\eta=0$ to the GOE distribution for $\eta>0$ more carefully and study the numerical distributions at very small values of $\eta$. Fig.~{\ref{fig:distributions_small_eta}} shows the distribution of level spacing for four different values of $\eta$ for a finite system of size $L=12$, at quarter filling for $U/J=2$.
The main difference between the Poisson distribution and the GOE like distribution is the finite value arising in the Poisson distribution at $r=0$ and its suppression due to the level repulsion in the GOE like ensemble. This shows that focusing on the manifestation
of level repulsion at small values of $r$ is much more sensitive than merely tracking the mean value of $r$. 
In order to resolve the steep distributions at low values or $r$ we choose a small bin size to represent the distributions. Already at very small values of $\eta/J=0.001,0.01$ a clear suppression of the value in the first bins is visible for the considered chain length. Whereas for $\eta/J=0.001$ this is only the case for the first bin, at $\eta/J=0.01$ already about a handful of bins contributes to the decrease of the distribution at low values of $r$ due to the level repulsion. For larger values of $\eta/J=0.05$ and $\eta/J=0.1$, the vanishing distribution at $r=0$ is clearly visible and the distribution approaches the GOE distribution.

The GOE distribution is expected for any finite value of the alternating potential ($\eta/J\neq0$) for sufficiently large system sizes. We see in Fig.~\ref{fig:different_sizes} that the suppression of the distribution at low values of $r$ is very sensible to system size. Whereas for $L=12$ still the distribution for low $r$ lies considerably above the GOE distribution, this deviation is already reduced for $L=16$ and the distribution lies much closer to the GOE distribution. This hints that already at small values of the alternating potential the level distribution follows the predictions from the GOE if sufficiently large system sizes are considered.

Also for large values of approximately $\eta/J \gtrsim 2$ the numerically found average value of $r$ drops below the value of the GOE ensemble (cf.~Fig.~\ref{fig:r_varying_eta}). We attribute the deviations from the expected GOE ensemble to finite size effects as discussed for small values of $\eta/J$. 
This is supported by the results obtained for larger system sizes in Fig.~\ref{fig:r_varying_eta} moving towards the GOE predictions with increasing system size both at small and large $\eta/J$. 
The deviations between the different system lengths are in particular large for low and large values of the alternating potential, where the GOE expectation is not yet met. 

The dependence of the average value of $r$ on the interaction strength is shown in Fig.~\ref{fig:r_varying_U} for two different value of $\eta$. A drastic rise is clearly seen from the non-interacting integrable case at $U=0$ which lies close to the predictions for the Poisson distribution to a finite interaction strength. Already at interaction strength of about $U/J\approx 1$ an almost steady value is reached for $\eta/J=1$, which agrees well with the predictions for a GOE ensemble even for a smaller system size considered [see Fig.~\ref{fig:r_varying_U}(a)]. 
At larger value of the alternating potential, $\eta/J=2$ similar behavior is observed but a larger system size of $L=16$ is needed to reach the expected GOE value due to finite size effects as discussed above [see Fig.~\ref{fig:r_varying_U}(b)]. 
At large values of the interaction we would expect again that deviations from the GOE ensemble arise due to the large separations occurring in the energy scales.

\begin{figure}[h!]
\centering
\includegraphics[width=0.95\linewidth]{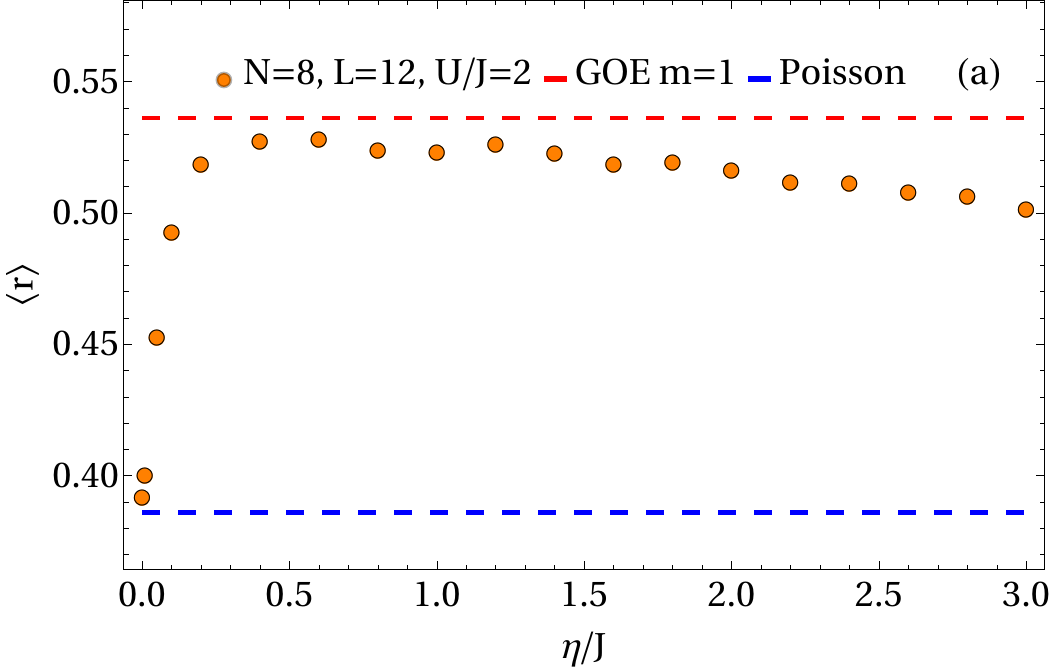}
\includegraphics[width=0.95\linewidth]{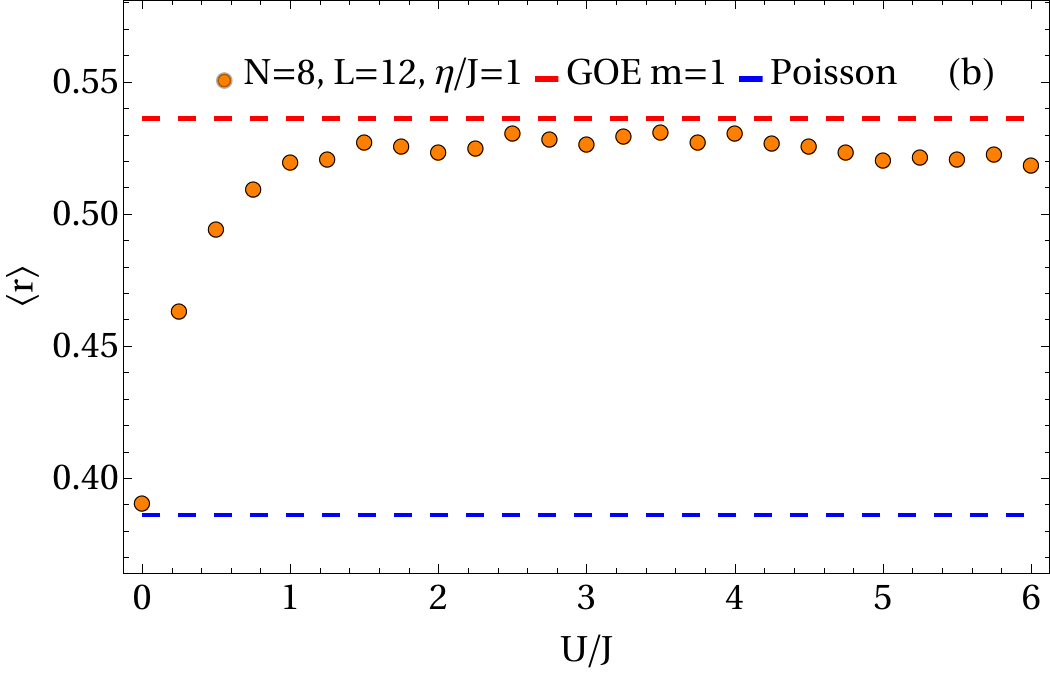}
\caption{Evolution of the mean value $\langle r \rangle$ of the ratios of consecutive level-spacings (a) with the alternating potential $\eta/J$ at $U/J=2$, and (b) with the interaction strength $U/J$ at $\eta/J=1$. The value for the Poissonian distribution and the GOE are marked by horizontal dashed lines. 
  For system size $N=8$ and $L=12$  combining $s=2,3,4$ and all $k,p$ for $m_z = 0$. 
 }
\label{fig:filling_2_3}
\end{figure}

In order to show that the discussed behavior is generic for different fillings, we show in Fig.~\ref{fig:filling_2_3} the dependence of the average value of $r$ on both the alternating potential $\eta$ and the interaction strength $U$ for $N/L=2/3$. We find that the shown behavior resembles very much the behavior of the quarter filling. The only difference is that the deviations from the GOE value at a larger values of the alternating potential are smaller than for quarter filling. 

We therefore conclude this section, that the energy spectra of the ionic Hubbard model at a generic filling  (here $N/L=1/2$ and $N/L=2/3$) nicely follow the expected behavior of a GOE ensemble and therefore point towards a chaotic nature of the model.

\subsection{Half filling}
\label{sec:half_filling}

\begin{figure}[h!]
\centering
\includegraphics[width=0.95\linewidth]{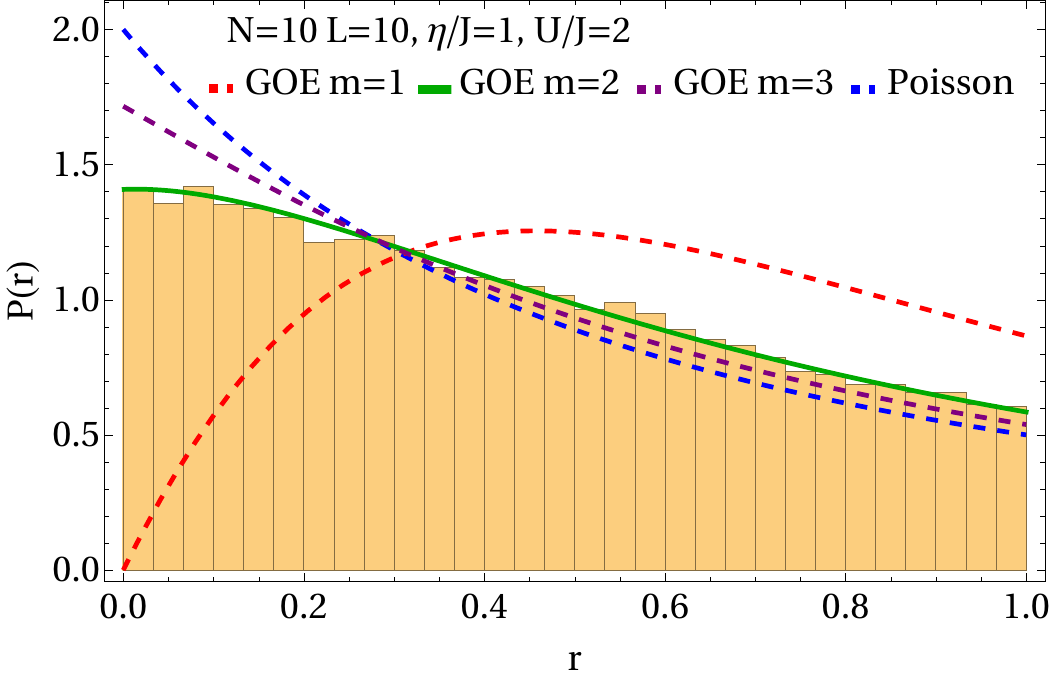}
\caption{Distribution of the ratios of consecutive level-spacings for $N = 10$, $L = 10$ for the symmetry sector $m_z = 0$ combining all $s$ and $k,p$ symmetry sectors. The parameters of the system are $U/J = 2$ and $\eta/J = 1$. The mean value is given by $\langle r \rangle = 
  0.426$.}
\label{fig:halffilling}
\end{figure}

\begin{figure}[h!]
  \centering
\includegraphics[width=0.95\linewidth]{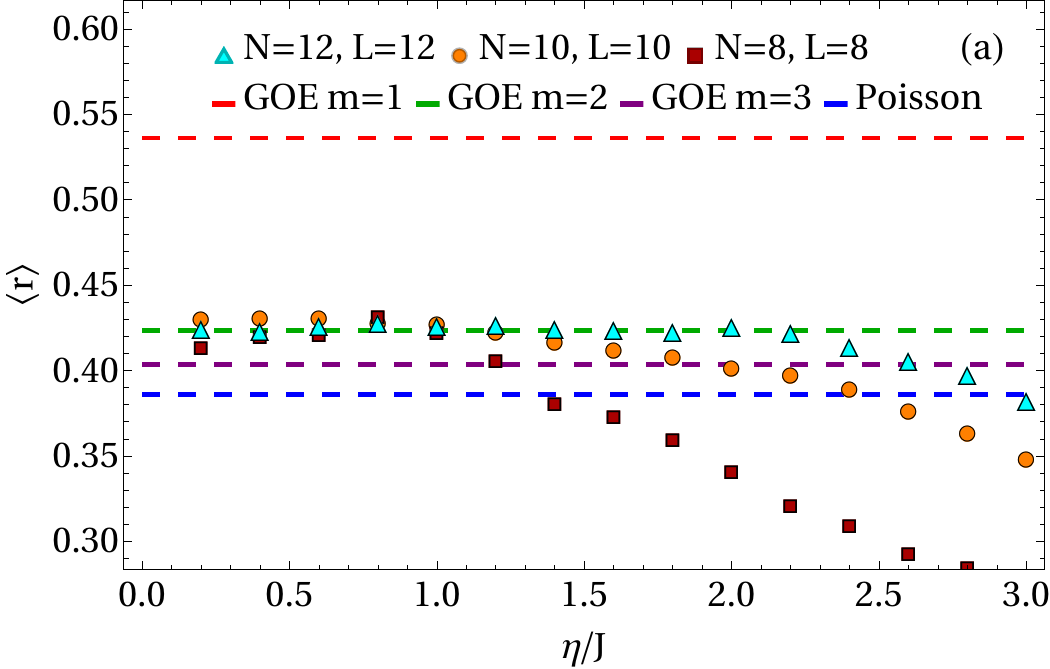}
\includegraphics[width=0.95\linewidth]{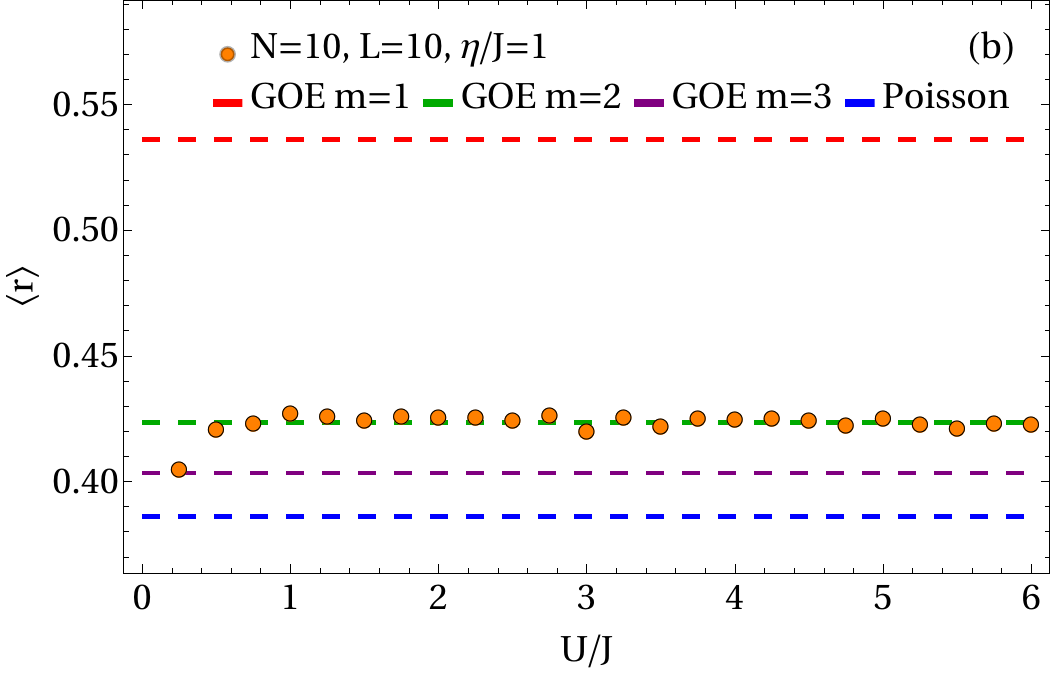}
\caption{Evolution of the mean value $\langle r \rangle$ of the ratios of consecutive level-spacings with (a) the alternating potential  at $U/J$=1 and (b) the interaction strength $U/J$ at $\eta/J=1$. The value for the Poissonian distribution and the GOE for different $m$ are marked by horizontal dashed lines (see Table~\ref{tab:values}).} 
\label{fig:half_r_eta_U}
\end{figure}

For the case of half filling $N/L=1$, the obtained distributions change drastically in comparison to what we described for quarter filling, as seen in Fig.~\ref{fig:halffilling}.
This motivated us to investigate this filling in more detail.
From our symmetry considerations, see Sec.~\ref{sec:sym}, we know that in the case of half filling an additional symmetry, Eq.~(\ref{eq:particle_hole}), is present in the ionic Hubbard model. As it was pointed out in Ref.~\cite{GiraudAlet2020} such an additional symmetry changes the behavior seen in the spectra drastically, if one does not consider the different symmetry blocks independently. 
As seen in Sec.~\ref{sec:statistics} depending on how many symmetries blocks are considered together, the distribution of the consecutive level spacing changes from the GOE like distribution to a more Poisson like distribution \cite{GiraudAlet2020}.
In Fig.~\ref{fig:halffilling} the distribution of the consecutive level spacings is shown for the case of half filling, where we have separated the spectrum into the blocks of the symmetries, Sec.~\ref{sec:sym} (a)-(d), but not into the additional particle hole symmetry, Sec.~\ref{sec:sym} (e). Note that resulting distribution neither follows a GOE like distribution nor a Poisson like distribution, but lies somewhere in between. These distributions are genuinely in between GOE and Poissonian, and do not converge to either of them with larger system sizes, unlike the crossover behavior discussed in the previous section. Using the predictions from Ref.~\cite{GiraudAlet2020} we confront the obtained distribution to the ones which correspond to the presence of two ($m=2$) or three ($m=3$) equally sized symmetry blocks (see Sec.~\ref{sec:statistics}). The numerical results lie very close to the results for a GOE RMT with {\em two} symmetry blocks present, as expected from our symmetry analysis.

This is further supported considering the mean value of $r$
and its dependence on the value of the alternating potential shown in Fig.~\ref{fig:half_r_eta_U} (a). As for the case of quarter filling, at $L=12$ intermediate value of $0.5\lesssim\eta/J\lesssim2$ lead to an excellent agreement with the predicted value of the one predicted for the distribution GOE $m=2$. Larger deviations occur at larger values of $\eta$. We attribute these again to finite size effects, as we can show that the longer system sizes considered strongly approach the expected mean value compared to the smaller system sizes. The average value of $r$ for the system size $L=12$ already agrees nicely up to $\eta/J\approx 2$ to the expected value fo the distribution of the GOE with $m=2$.  

In Fig.~\ref{fig:half_r_eta_U} (b) the average value and its dependence on different values of $U/J$ is shown. For $U/J>0.5$, the obtained main value fluctuates very closely around the one predicted for the distribution GOE $m=2$ and is distinct from the one predicted for the distribution with $m=3$ equal symmetry blocks. 
 
To conclude our investigation at half-filling, let us note, that the form of the distribution drastically changes and cannot be distinguished from the Poisson distribution, if the level statistics is not separated with respect to the symmetries, Sec.~\ref{sec:sym} (a)-(d), as shown in Appendix \ref{app:withoutS}.

Thus, our entire results indicate that the ionic Hubbard model shows similar properties as the GOE like ensemble, if all known symmetries are considered.

\subsection{Half filling - breaking the particle-hole symmetry}
\label{sec:half_filling_broken_sym}

\begin{figure}[ht!]
\centering
\includegraphics[width=0.95\linewidth]{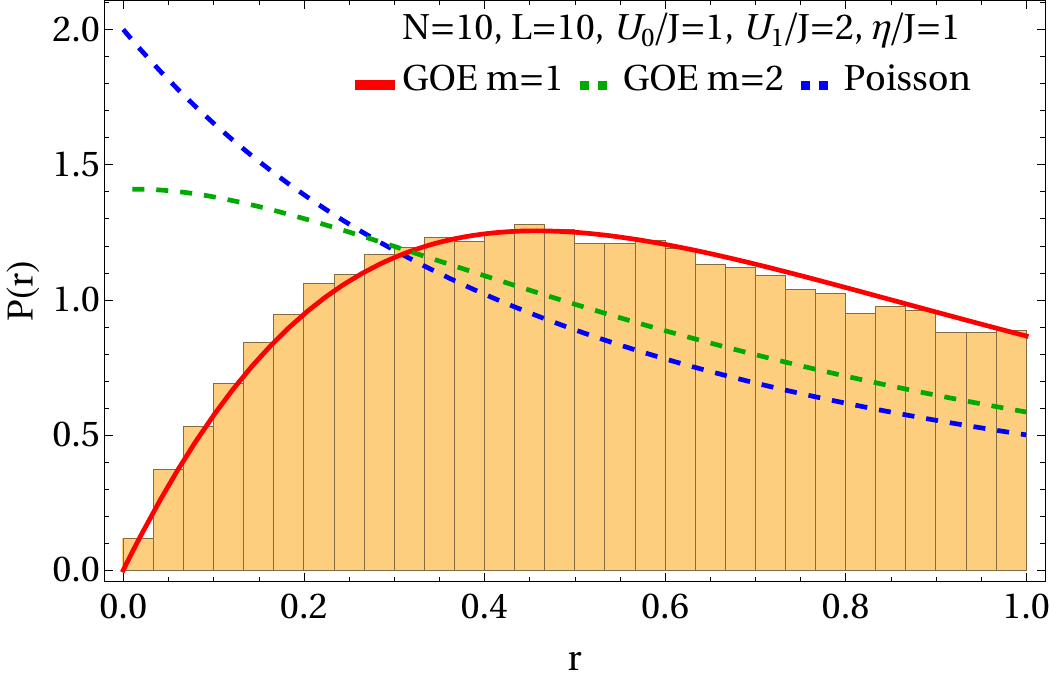}
\caption{Distribution of the ratios of consecutive level-spacings for $N = 10$, $L = 10$ for the symmetry sector $m_z = 0$ combining all $s$ and $k,p$ symmetry sectors. The parameters of the system are $U_0/J = 1$ $U_1/J = 2$ and $\eta/J = 1$, and 30 bins. The mean value is given by $\langle r \rangle = 0.527$.}
\label{fig:diffmodel}
\end{figure}

\begin{figure}[h!]
\centering
\includegraphics[width=0.95\linewidth]{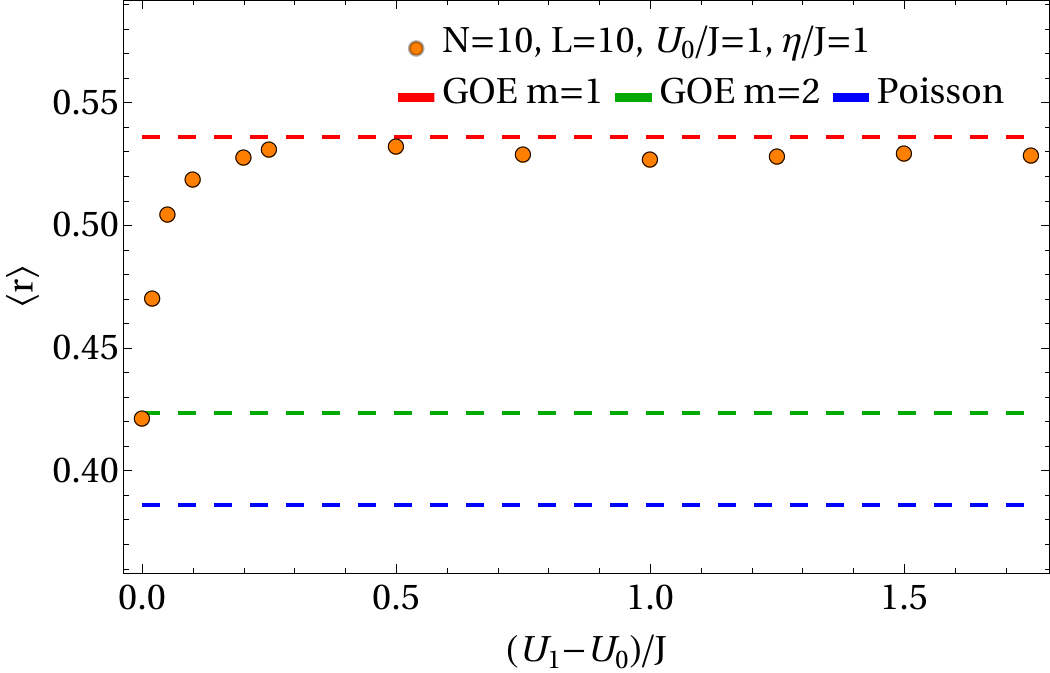}
\caption{Evolution of the mean value $\langle r \rangle$ of the ratios of consecutive level-spacings with the difference of the interaction strengths $(U_1-U_0)/J$ at $\eta/J=1$, $U_0/J=1$. The horizontal dashed lines mark the expected values for the Poisson distribution and for the GOE with different $m$ (see Table \ref{tab:values}).
  System size $N=10$, $L=10$ considered symmetry sector $m_z=0$ combining all $s$ and $k,p$ symmetry sectors.}
\label{fig:diffU}
\end{figure}

In order to further corroborate our finding of the importance of the particle-hole like symmetry at half filling, we introduce an additional term in the Hamiltonian which explicitly breaks this symmetry. This is achieved by substituting the homogeneous on-site interaction with an interaction using an additional alternation in the on-site interaction term:
\begin{equation}
    H_{\text{int}} = U_0 \sum_{j = 1}^{L/2} n_{2j,\uparrow} n_{2j,\downarrow} + U_1 \sum_{j = 1}^{L/2} n_{2j+1,\uparrow} n_{2j+1,\downarrow}
\end{equation}
This term breaks the particle-hole like symmetry, but leaves all the other discussed symmetries (a)-(d), such as the reflection, translation and spin rotation symmetry invariant. As shown in Fig.~\ref{fig:diffmodel} the presence of such a symmetry breaking term leads to a distribution which follows very clearly the GOE distribution for a {\em single} ($m=1$) symmetry block. 
In order to investigate this behavior in more detail we show in Fig.~\ref{fig:diffU} the average value of $r$ with the symmetry breaking term,~i.e. with the difference in the interaction strength $U_1-U_0$. A very steep rise of the average value of $r$ is seen at small amplitudes of the symmetry breaking term, but already at $(U_1-U_0)/J \approx 0.2$, the average value stabilizes close to the expected value of the GOE $m=1$ ensemble. We expect that this rise becomes more and more steep with increasing system size signaling a direct change of the behavior from the $m=2$ situation for the presence of the particle-hole like symmetry to the standard GOE ($m=1$) behavior when this additional symmetry is
explicitly broken. 

\section{Conclusion}
In this work, we have investigated the energy level statistics of the ionic Hubbard model and a generalization which breaks the particle-hole like symmetry at half filling. We find that in general the ionic Hubbard model and its generalization show, when all known symmetries are considered, the spectral features of a GOE ensemble. This implies the chaotic behavior of the level statistics of the ionic Hubbard model. Following the conjecture that the GOE like statistics indicates non-integrability of a many-body Hamiltonian \cite{PoilblancMontambaux1993}, to which to our knowledge no generic counterexamples are known, our results indicate the non-integrability of the ionic Hubbard model and its generalization. Our findings are in agreement with the general belief that the ionic Hubbard model is, for generic parameters, a non-integrable model.

\section*{Acknowledgments}
We thank F. Essler, T. Giamarchi, and A. Rosch for stimulating discussions.  AML acknowledges T.~Busse and M.~Haque for discussions on random matrices. We acknowledge funding from the Deutsche Forschungsgemeinschaft (DFG, German Research Foundation) in particular under project number 277625399 - TRR 185 (B3,B4) and project number 277146847 - CRC 1238 (C05) and under Germany’s Excellence Strategy – Cluster of Excellence Matter and Light for Quantum Computing (ML4Q) EXC 2004/1 – 390534769 and the European Research Council (ERC) under the Horizon 2020 research and innovation programme, grant agreement No.~648166 (Phonton).
This work was supported in part by the Swiss National Science Foundation under Division II.

\renewcommand{\theequation}{A\arabic{equation}}
\renewcommand{\thefigure}{A\arabic{figure}}
\setcounter{equation}{0}
\setcounter{figure}{0}

\appendix
\section{Importance of symmetries in the numerical analysis of the spectral properties}
\label{app:withoutS}

\begin{figure}[h!]
	\centering
	\includegraphics[width=0.95\linewidth]{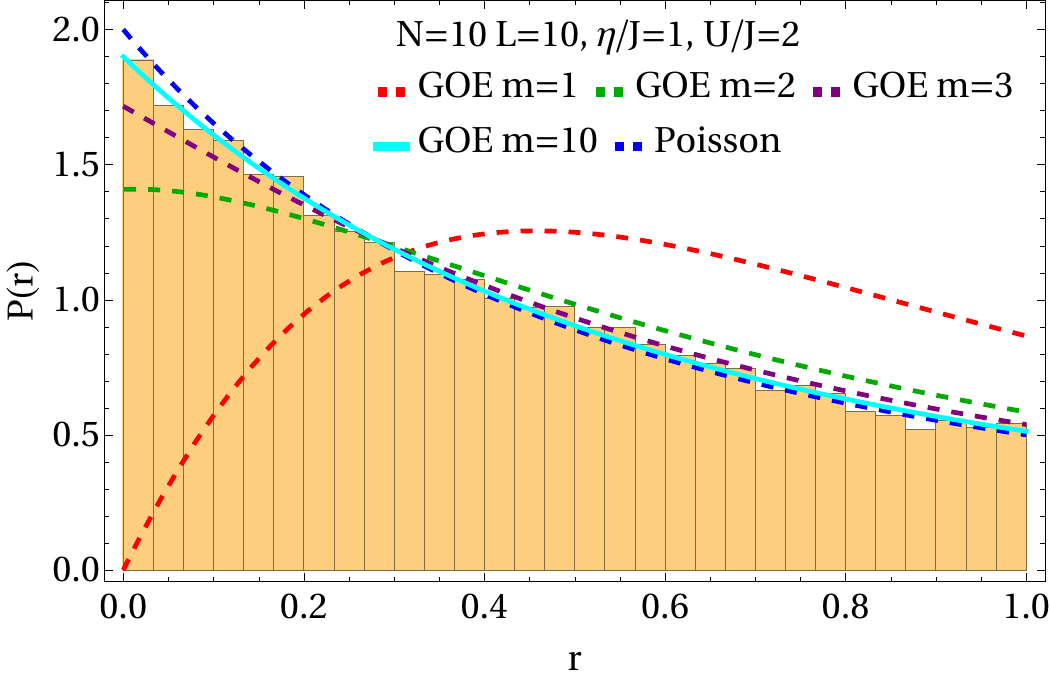}
	\caption{Distribution of the ratios of consecutive level-spacings for $N = 10$, $L = 10$ for the symmetry sector $m_z = 0$, but not separating for the $s$ quantum number and combining data from all $k,p$ sectors. The parameters of the system are $U/J = 2$  and $\eta/J= 1$. The mean value is given by $\langle r \rangle = 0.392$.}
	\label{int_withoutS}
\end{figure}

In this appendix, we show how neglecting the separation into the different symmetry sectors the in principle GOE like distribution of the consecutive level spacings approaches the Poisson-like distribution.
In particular, we focus on half filling, for parameters close to the ones considered in Hosseinzadeh et al. \cite{JafariHosseinzadeh2020} where the Poisson statistics was identified.

To emphasize the effects of the separation into the different symmetry sectors, we present in Fig.~\ref{int_withoutS} the distribution of the consecutive level spacings at half filling, but only taking a fixed $S_z$ sector, and not a fixed sector of the total spin $S$. 
In this case the expected distribution for a chaotic system would be a GOE with $m=10$ symmetry blocks \cite{GiraudAlet2020}. The number of blocks results from the five symmetry sectors due to the total spin, which are of unequal size, which should furthermore be split in half due to the particle-hole symmetry present at half filling (see Sec.~\ref{sec:half_filling}).
The sizes of the symmetry sectors are: $s=0$, 3880 states; $s=1$, 5940 states; $s=2$, 2475 states; $s=3$, 385 states; $s=4$, 20 states.
We can notice in Fig.~\ref{int_withoutS} that this GOE distribution with $m=10$ is very close to a Poisson distribution due to the large number of symmetry blocks. Our numerical data is in agreement with this distribution and thus, also lies very close to the Poisson distribution.  
This again underlines the importance of the separation of the spectra into the different symmetry sectors.  

\renewcommand{\theequation}{B\arabic{equation}}
\renewcommand{\thefigure}{B\arabic{figure}}
\setcounter{equation}{0}
\setcounter{figure}{0}

\section{The implementation of the spin rotation symmetries}\label{app:impl_symm}

As explained in the main text the quantum numbers associated to the spin rotation are the total spin $S$ with quantum number $s$ and the spin in $z$-direction $S^z$ with quantum number $m_z$ (which varies from $-s$ to $s$) with the corresponding eigenvector in each symmetry sector labeled as $\vert s,m_z\rangle$. In our computation only the quantum number of $S^z$ is implemented but one can take the advantage of the degeneracy of the eigenvalues for a fixed total spin $s$ and different $m_z$'s. The eigenvectors $\vert s,m_z\rangle$ and $\vert s,m_z'\rangle$ are connected to each other with the ladder operators  $S^+$ and $S^-$. As the Hamiltonian commutes with the ladder operators, the eigenvectors $\vert s,m_z\rangle$ and $\vert s,m_z'\rangle$ are degenerate. Thus, knowing all eigenvalues for different $m_z$'s one can pick the ones for the desired quantum numbers $s$. We start with the symmetry block with largest possible value of $m_z$ (for $N$ spin-half particles this value is $N/2$) which corresponds only to one total spin $s=m_z$ and the solution is clear. In the next symmetry sector with smaller quantum number $m_z=s-1$ there exist the eigenvalues for two quantum numbers, $s$ and $s-1$, the ones with total spin $s$ are the degenerate ones with the previous sector we looked at and the remaining eigenvalues correspond to the symmetry sector with total spin $s-1$. One can decrease the quantum number $m_z$ to zero and and with this procedure find all eigenvalues labeled with both quantum numbers $s$ and $m_z$. Note, that the eigenvalues corresponding to the negative spin in $z$-direction ($m_z<0$ ) for a fixed total spin $s$ are degenerate with the ones with its positive value ($\vert m_z\vert$).


\begin{thebibliography}{70}%
\makeatletter
\providecommand \@ifxundefined [1]{%
 \@ifx{#1\undefined}
}%
\providecommand \@ifnum [1]{%
 \ifnum #1\expandafter \@firstoftwo
 \else \expandafter \@secondoftwo
 \fi
}%
\providecommand \@ifx [1]{%
 \ifx #1\expandafter \@firstoftwo
 \else \expandafter \@secondoftwo
 \fi
}%
\providecommand \natexlab [1]{#1}%
\providecommand \enquote  [1]{``#1''}%
\providecommand \bibnamefont  [1]{#1}%
\providecommand \bibfnamefont [1]{#1}%
\providecommand \citenamefont [1]{#1}%
\providecommand \href@noop [0]{\@secondoftwo}%
\providecommand \href [0]{\begingroup \@sanitize@url \@href}%
\providecommand \@href[1]{\@@startlink{#1}\@@href}%
\providecommand \@@href[1]{\endgroup#1\@@endlink}%
\providecommand \@sanitize@url [0]{\catcode `\\12\catcode `\$12\catcode
  `\&12\catcode `\#12\catcode `\^12\catcode `\_12\catcode `\%12\relax}%
\providecommand \@@startlink[1]{}%
\providecommand \@@endlink[0]{}%
\providecommand \url  [0]{\begingroup\@sanitize@url \@url }%
\providecommand \@url [1]{\endgroup\@href {#1}{\urlprefix }}%
\providecommand \urlprefix  [0]{URL }%
\providecommand \Eprint [0]{\href }%
\providecommand \doibase [0]{http://dx.doi.org/}%
\providecommand \selectlanguage [0]{\@gobble}%
\providecommand \bibinfo  [0]{\@secondoftwo}%
\providecommand \bibfield  [0]{\@secondoftwo}%
\providecommand \translation [1]{[#1]}%
\providecommand \BibitemOpen [0]{}%
\providecommand \bibitemStop [0]{}%
\providecommand \bibitemNoStop [0]{.\EOS\space}%
\providecommand \EOS [0]{\spacefactor3000\relax}%
\providecommand \BibitemShut  [1]{\csname bibitem#1\endcsname}%
\let\auto@bib@innerbib\@empty
\bibitem [{\citenamefont {Srednicki}(1994)}]{Srednicki1994}%
  \BibitemOpen
  \bibfield  {author} {\bibinfo {author} {\bibfnamefont {M.}~\bibnamefont
  {Srednicki}},\ }\emph {Chaos and quantum thermalization},\ \href {\doibase
  10.1103/PhysRevE.50.888} {\bibfield  {journal} {\bibinfo  {journal} {Phys.
  Rev. E}\ }\textbf {\bibinfo {volume} {50}},\ \bibinfo {pages} {888} (\bibinfo
  {year} {1994})}\BibitemShut {NoStop}%
\bibitem [{\citenamefont {Kollath}\ \emph {et~al.}(2007)\citenamefont
  {Kollath}, \citenamefont {L\"auchli},\ and\ \citenamefont
  {Altman}}]{KollathAltman2007}%
  \BibitemOpen
  \bibfield  {author} {\bibinfo {author} {\bibfnamefont {C.}~\bibnamefont
  {Kollath}}, \bibinfo {author} {\bibfnamefont {A.~M.}\ \bibnamefont
  {L\"auchli}}, \ and\ \bibinfo {author} {\bibfnamefont {E.}~\bibnamefont
  {Altman}},\ }\emph {Quench Dynamics and Nonequilibrium Phase Diagram of the
  Bose-Hubbard Model},\ \href {\doibase 10.1103/PhysRevLett.98.180601}
  {\bibfield  {journal} {\bibinfo  {journal} {Phys. Rev. Lett.}\ }\textbf
  {\bibinfo {volume} {98}},\ \bibinfo {pages} {180601} (\bibinfo {year}
  {2007})}\BibitemShut {NoStop}%
\bibitem [{\citenamefont {Polkovnikov}\ \emph {et~al.}(2011)\citenamefont
  {Polkovnikov}, \citenamefont {Sengupta}, \citenamefont {Silva},\ and\
  \citenamefont {Vengalattore}}]{PolkovnikovVengalattore2011}%
  \BibitemOpen
  \bibfield  {author} {\bibinfo {author} {\bibfnamefont {A.}~\bibnamefont
  {Polkovnikov}}, \bibinfo {author} {\bibfnamefont {K.}~\bibnamefont
  {Sengupta}}, \bibinfo {author} {\bibfnamefont {A.}~\bibnamefont {Silva}}, \
  and\ \bibinfo {author} {\bibfnamefont {M.}~\bibnamefont {Vengalattore}},\
  }\emph {Colloquium: Nonequilibrium dynamics of closed interacting quantum
  systems},\ \href {\doibase 10.1103/RevModPhys.83.863} {\bibfield  {journal}
  {\bibinfo  {journal} {Rev. Mod. Phys.}\ }\textbf {\bibinfo {volume} {83}},\
  \bibinfo {pages} {863} (\bibinfo {year} {2011})}\BibitemShut {NoStop}%
\bibitem [{\citenamefont {Nandkishore}\ and\ \citenamefont
  {Huse}(2015)}]{NandkishoreHuse2015}%
  \BibitemOpen
  \bibfield  {author} {\bibinfo {author} {\bibfnamefont {R.}~\bibnamefont
  {Nandkishore}}\ and\ \bibinfo {author} {\bibfnamefont {D.~A.}\ \bibnamefont
  {Huse}},\ }\emph {Many-Body Localization and Thermalization in Quantum
  Statistical Mechanics},\ \href {\doibase
  10.1146/annurev-conmatphys-031214-014726} {\bibfield  {journal} {\bibinfo
  {journal} {Annual Review of Condensed Matter Physics}\ }\textbf {\bibinfo
  {volume} {6}},\ \bibinfo {pages} {15} (\bibinfo {year} {2015})}\BibitemShut
  {NoStop}%
\bibitem [{\citenamefont {Abanin}\ \emph {et~al.}(2019)\citenamefont {Abanin},
  \citenamefont {Altman}, \citenamefont {Bloch},\ and\ \citenamefont
  {Serbyn}}]{AbaninSerbyn2019}%
  \BibitemOpen
  \bibfield  {author} {\bibinfo {author} {\bibfnamefont {D.~A.}\ \bibnamefont
  {Abanin}}, \bibinfo {author} {\bibfnamefont {E.}~\bibnamefont {Altman}},
  \bibinfo {author} {\bibfnamefont {I.}~\bibnamefont {Bloch}}, \ and\ \bibinfo
  {author} {\bibfnamefont {M.}~\bibnamefont {Serbyn}},\ }\emph {Colloquium:
  Many-body localization, thermalization, and entanglement},\ \href {\doibase
  10.1103/RevModPhys.91.021001} {\bibfield  {journal} {\bibinfo  {journal}
  {Rev. Mod. Phys.}\ }\textbf {\bibinfo {volume} {91}},\ \bibinfo {pages}
  {021001} (\bibinfo {year} {2019})}\BibitemShut {NoStop}%
\bibitem [{\citenamefont {Haake}(2000)}]{Haake2000}%
  \BibitemOpen
  \bibfield  {author} {\bibinfo {author} {\bibfnamefont {F.}~\bibnamefont
  {Haake}},\ }\href {\doibase 10.1007/978-3-642-05428-0} {\emph {\bibinfo
  {title} {Quantum Signatures of Chaos}}}\ (\bibinfo  {publisher} {Springer,
  Berlin Heidelberg New York},\ \bibinfo {year} {2000})\BibitemShut {NoStop}%
\bibitem [{\citenamefont {Basko}\ \emph {et~al.}(2006)\citenamefont {Basko},
  \citenamefont {Aleiner},\ and\ \citenamefont
  {Altshuler}}]{BaskoAltshuler2006}%
  \BibitemOpen
  \bibfield  {author} {\bibinfo {author} {\bibfnamefont {D.}~\bibnamefont
  {Basko}}, \bibinfo {author} {\bibfnamefont {I.}~\bibnamefont {Aleiner}}, \
  and\ \bibinfo {author} {\bibfnamefont {B.}~\bibnamefont {Altshuler}},\ }\emph
  {Metal–insulator transition in a weakly interacting many-electron system
  with localized single-particle states},\ \href {\doibase
  https://doi.org/10.1016/j.aop.2005.11.014} {\bibfield  {journal} {\bibinfo
  {journal} {Annals of Physics}\ }\textbf {\bibinfo {volume} {321}},\ \bibinfo
  {pages} {1126} (\bibinfo {year} {2006})}\BibitemShut {NoStop}%
\bibitem [{\citenamefont {Gornyi}\ \emph {et~al.}(2005)\citenamefont {Gornyi},
  \citenamefont {Mirlin},\ and\ \citenamefont {Polyakov}}]{GornyiPolyakov2005}%
  \BibitemOpen
  \bibfield  {author} {\bibinfo {author} {\bibfnamefont {I.~V.}\ \bibnamefont
  {Gornyi}}, \bibinfo {author} {\bibfnamefont {A.~D.}\ \bibnamefont {Mirlin}},
  \ and\ \bibinfo {author} {\bibfnamefont {D.~G.}\ \bibnamefont {Polyakov}},\
  }\emph {Interacting Electrons in Disordered Wires: Anderson Localization and
  Low-$T$ Transport},\ \href {\doibase 10.1103/PhysRevLett.95.206603}
  {\bibfield  {journal} {\bibinfo  {journal} {Phys. Rev. Lett.}\ }\textbf
  {\bibinfo {volume} {95}},\ \bibinfo {pages} {206603} (\bibinfo {year}
  {2005})}\BibitemShut {NoStop}%
\bibitem [{\citenamefont {Pal}\ and\ \citenamefont {Huse}(2010)}]{PalHuse2010}%
  \BibitemOpen
  \bibfield  {author} {\bibinfo {author} {\bibfnamefont {A.}~\bibnamefont
  {Pal}}\ and\ \bibinfo {author} {\bibfnamefont {D.~A.}\ \bibnamefont {Huse}},\
  }\emph {Many-body localization phase transition},\ \href {\doibase
  10.1103/PhysRevB.82.174411} {\bibfield  {journal} {\bibinfo  {journal} {Phys.
  Rev. B}\ }\textbf {\bibinfo {volume} {82}},\ \bibinfo {pages} {174411}
  (\bibinfo {year} {2010})}\BibitemShut {NoStop}%
\bibitem [{\citenamefont {Sala}\ \emph {et~al.}(2020)\citenamefont {Sala},
  \citenamefont {Rakovszky}, \citenamefont {Verresen}, \citenamefont {Knap},\
  and\ \citenamefont {Pollmann}}]{SalaPollmann2020}%
  \BibitemOpen
  \bibfield  {author} {\bibinfo {author} {\bibfnamefont {P.}~\bibnamefont
  {Sala}}, \bibinfo {author} {\bibfnamefont {T.}~\bibnamefont {Rakovszky}},
  \bibinfo {author} {\bibfnamefont {R.}~\bibnamefont {Verresen}}, \bibinfo
  {author} {\bibfnamefont {M.}~\bibnamefont {Knap}}, \ and\ \bibinfo {author}
  {\bibfnamefont {F.}~\bibnamefont {Pollmann}},\ }\emph {Ergodicity Breaking
  Arising from Hilbert Space Fragmentation in Dipole-Conserving Hamiltonians},\
  \href {\doibase 10.1103/PhysRevX.10.011047} {\bibfield  {journal} {\bibinfo
  {journal} {Phys. Rev. X}\ }\textbf {\bibinfo {volume} {10}},\ \bibinfo
  {pages} {011047} (\bibinfo {year} {2020})}\BibitemShut {NoStop}%
\bibitem [{\citenamefont {Khemani}\ \emph {et~al.}(2020)\citenamefont
  {Khemani}, \citenamefont {Hermele},\ and\ \citenamefont
  {Nandkishore}}]{KhemaniNandkishore2020}%
  \BibitemOpen
  \bibfield  {author} {\bibinfo {author} {\bibfnamefont {V.}~\bibnamefont
  {Khemani}}, \bibinfo {author} {\bibfnamefont {M.}~\bibnamefont {Hermele}}, \
  and\ \bibinfo {author} {\bibfnamefont {R.}~\bibnamefont {Nandkishore}},\
  }\emph {Localization from Hilbert space shattering: From theory to physical
  realizations},\ \href {\doibase 10.1103/PhysRevB.101.174204} {\bibfield
  {journal} {\bibinfo  {journal} {Phys. Rev. B}\ }\textbf {\bibinfo {volume}
  {101}},\ \bibinfo {pages} {174204} (\bibinfo {year} {2020})}\BibitemShut
  {NoStop}%
\bibitem [{\citenamefont {Shiraishi}\ and\ \citenamefont
  {Mori}(2017)}]{ShiraishiMori2017}%
  \BibitemOpen
  \bibfield  {author} {\bibinfo {author} {\bibfnamefont {N.}~\bibnamefont
  {Shiraishi}}\ and\ \bibinfo {author} {\bibfnamefont {T.}~\bibnamefont
  {Mori}},\ }\emph {Systematic Construction of Counterexamples to the
  Eigenstate Thermalization Hypothesis},\ \href {\doibase
  10.1103/PhysRevLett.119.030601} {\bibfield  {journal} {\bibinfo  {journal}
  {Phys. Rev. Lett.}\ }\textbf {\bibinfo {volume} {119}},\ \bibinfo {pages}
  {030601} (\bibinfo {year} {2017})}\BibitemShut {NoStop}%
\bibitem [{\citenamefont {Turner}\ \emph {et~al.}(2018)\citenamefont {Turner},
  \citenamefont {Michailidis}, \citenamefont {Abanin}, \citenamefont {Serbyn},\
  and\ \citenamefont {Papi{\'{c}}}}]{TurnerPapic2018}%
  \BibitemOpen
  \bibfield  {author} {\bibinfo {author} {\bibfnamefont {C.~J.}\ \bibnamefont
  {Turner}}, \bibinfo {author} {\bibfnamefont {A.~A.}\ \bibnamefont
  {Michailidis}}, \bibinfo {author} {\bibfnamefont {D.~A.}\ \bibnamefont
  {Abanin}}, \bibinfo {author} {\bibfnamefont {M.}~\bibnamefont {Serbyn}}, \
  and\ \bibinfo {author} {\bibfnamefont {Z.}~\bibnamefont {Papi{\'{c}}}},\
  }\emph {Weak ergodicity breaking from quantum many-body scars},\ \href
  {\doibase 10.1038/s41567-018-0137-5} {\bibfield  {journal} {\bibinfo
  {journal} {Nature Physics}\ }\textbf {\bibinfo {volume} {14}},\ \bibinfo
  {pages} {745} (\bibinfo {year} {2018})}\BibitemShut {NoStop}%
\bibitem [{\citenamefont {Khemani}\ \emph {et~al.}(2019)\citenamefont
  {Khemani}, \citenamefont {Laumann},\ and\ \citenamefont
  {Chandran}}]{KhemaniChandran2019}%
  \BibitemOpen
  \bibfield  {author} {\bibinfo {author} {\bibfnamefont {V.}~\bibnamefont
  {Khemani}}, \bibinfo {author} {\bibfnamefont {C.~R.}\ \bibnamefont
  {Laumann}}, \ and\ \bibinfo {author} {\bibfnamefont {A.}~\bibnamefont
  {Chandran}},\ }\emph {Signatures of integrability in the dynamics of
  Rydberg-blockaded chains},\ \href {\doibase 10.1103/PhysRevB.99.161101}
  {\bibfield  {journal} {\bibinfo  {journal} {Phys. Rev. B}\ }\textbf {\bibinfo
  {volume} {99}},\ \bibinfo {pages} {161101} (\bibinfo {year}
  {2019})}\BibitemShut {NoStop}%
\bibitem [{\citenamefont {Ho}\ \emph {et~al.}(2019)\citenamefont {Ho},
  \citenamefont {Choi}, \citenamefont {Pichler},\ and\ \citenamefont
  {Lukin}}]{HoLukin2019}%
  \BibitemOpen
  \bibfield  {author} {\bibinfo {author} {\bibfnamefont {W.~W.}\ \bibnamefont
  {Ho}}, \bibinfo {author} {\bibfnamefont {S.}~\bibnamefont {Choi}}, \bibinfo
  {author} {\bibfnamefont {H.}~\bibnamefont {Pichler}}, \ and\ \bibinfo
  {author} {\bibfnamefont {M.~D.}\ \bibnamefont {Lukin}},\ }\emph {Periodic
  Orbits, Entanglement, and Quantum Many-Body Scars in Constrained Models:
  Matrix Product State Approach},\ \href {\doibase
  10.1103/PhysRevLett.122.040603} {\bibfield  {journal} {\bibinfo  {journal}
  {Phys. Rev. Lett.}\ }\textbf {\bibinfo {volume} {122}},\ \bibinfo {pages}
  {040603} (\bibinfo {year} {2019})}\BibitemShut {NoStop}%
\bibitem [{\citenamefont {Bernien}\ \emph {et~al.}(2017)\citenamefont
  {Bernien}, \citenamefont {Schwartz}, \citenamefont {Keesling}, \citenamefont
  {Levine}, \citenamefont {Omran}, \citenamefont {Pichler}, \citenamefont
  {Choi}, \citenamefont {Zibrov}, \citenamefont {Endres}, \citenamefont
  {Greiner}, \citenamefont {Vuleti{\'{c}}},\ and\ \citenamefont
  {Lukin}}]{BernienLukin2017}%
  \BibitemOpen
  \bibfield  {author} {\bibinfo {author} {\bibfnamefont {H.}~\bibnamefont
  {Bernien}}, \bibinfo {author} {\bibfnamefont {S.}~\bibnamefont {Schwartz}},
  \bibinfo {author} {\bibfnamefont {A.}~\bibnamefont {Keesling}}, \bibinfo
  {author} {\bibfnamefont {H.}~\bibnamefont {Levine}}, \bibinfo {author}
  {\bibfnamefont {A.}~\bibnamefont {Omran}}, \bibinfo {author} {\bibfnamefont
  {H.}~\bibnamefont {Pichler}}, \bibinfo {author} {\bibfnamefont
  {S.}~\bibnamefont {Choi}}, \bibinfo {author} {\bibfnamefont {A.~S.}\
  \bibnamefont {Zibrov}}, \bibinfo {author} {\bibfnamefont {M.}~\bibnamefont
  {Endres}}, \bibinfo {author} {\bibfnamefont {M.}~\bibnamefont {Greiner}},
  \bibinfo {author} {\bibfnamefont {V.}~\bibnamefont {Vuleti{\'{c}}}}, \ and\
  \bibinfo {author} {\bibfnamefont {M.~D.}\ \bibnamefont {Lukin}},\ }\emph
  {Probing many-body dynamics on a 51-atom quantum simulator},\ \href {\doibase
  10.1038/nature24622} {\bibfield  {journal} {\bibinfo  {journal} {Nature}\
  }\textbf {\bibinfo {volume} {551}},\ \bibinfo {pages} {579} (\bibinfo {year}
  {2017})}\BibitemShut {NoStop}%
\bibitem [{\citenamefont {Mehta}(1991)}]{Mehtabook}%
  \BibitemOpen
  \bibfield  {author} {\bibinfo {author} {\bibfnamefont {M.~L.}\ \bibnamefont
  {Mehta}},\ }\href {\doibase https://doi.org/10.1016/C2009-0-22297-5} {\emph
  {\bibinfo {title} {Random Matrices}}},\ \bibinfo {edition} {2nd}\ ed.\
  (\bibinfo  {publisher} {Academic Press},\ \bibinfo {year} {1991})\BibitemShut
  {NoStop}%
\bibitem [{\citenamefont {Berry}\ \emph {et~al.}(1977)\citenamefont {Berry},
  \citenamefont {Tabor},\ and\ \citenamefont {Ziman}}]{BerryZiman1977}%
  \BibitemOpen
  \bibfield  {author} {\bibinfo {author} {\bibfnamefont {M.~V.}\ \bibnamefont
  {Berry}}, \bibinfo {author} {\bibfnamefont {M.}~\bibnamefont {Tabor}}, \ and\
  \bibinfo {author} {\bibfnamefont {J.~M.}\ \bibnamefont {Ziman}},\ }\emph
  {Level clustering in the regular spectrum},\ \href {\doibase
  10.1098/rspa.1977.0140} {\bibfield  {journal} {\bibinfo  {journal}
  {Proceedings of the Royal Society of London. A. Mathematical and Physical
  Sciences}\ }\textbf {\bibinfo {volume} {356}},\ \bibinfo {pages} {375}
  (\bibinfo {year} {1977})}\BibitemShut {NoStop}%
\bibitem [{\citenamefont {Bohigas}\ \emph {et~al.}(1984)\citenamefont
  {Bohigas}, \citenamefont {Giannoni},\ and\ \citenamefont
  {Schmit}}]{BohigasSchmit1984}%
  \BibitemOpen
  \bibfield  {author} {\bibinfo {author} {\bibfnamefont {O.}~\bibnamefont
  {Bohigas}}, \bibinfo {author} {\bibfnamefont {M.~J.}\ \bibnamefont
  {Giannoni}}, \ and\ \bibinfo {author} {\bibfnamefont {C.}~\bibnamefont
  {Schmit}},\ }\emph {Characterization of Chaotic Quantum Spectra and
  Universality of Level Fluctuation Laws},\ \href {\doibase
  10.1103/PhysRevLett.52.1} {\bibfield  {journal} {\bibinfo  {journal} {Phys.
  Rev. Lett.}\ }\textbf {\bibinfo {volume} {52}},\ \bibinfo {pages} {1}
  (\bibinfo {year} {1984})}\BibitemShut {NoStop}%
\bibitem [{\citenamefont {Poilblanc}\ \emph {et~al.}(1993)\citenamefont
  {Poilblanc}, \citenamefont {Ziman}, \citenamefont {Bellissard}, \citenamefont
  {Mila},\ and\ \citenamefont {Montambaux}}]{PoilblancMontambaux1993}%
  \BibitemOpen
  \bibfield  {author} {\bibinfo {author} {\bibfnamefont {D.}~\bibnamefont
  {Poilblanc}}, \bibinfo {author} {\bibfnamefont {T.}~\bibnamefont {Ziman}},
  \bibinfo {author} {\bibfnamefont {J.}~\bibnamefont {Bellissard}}, \bibinfo
  {author} {\bibfnamefont {F.}~\bibnamefont {Mila}}, \ and\ \bibinfo {author}
  {\bibfnamefont {G.}~\bibnamefont {Montambaux}},\ }\emph {Poisson vs. {GOE}
  Statistics in Integrable and Non-Integrable Quantum Hamiltonians},\ \href
  {\doibase 10.1209/0295-5075/22/7/010} {\bibfield  {journal} {\bibinfo
  {journal} {Europhysics Letters ({EPL})}\ }\textbf {\bibinfo {volume} {22}},\
  \bibinfo {pages} {537} (\bibinfo {year} {1993})}\BibitemShut {NoStop}%
\bibitem [{\citenamefont {Hosseinzadeh}\ and\ \citenamefont
  {Jafari}(2020)}]{JafariHosseinzadeh2020}%
  \BibitemOpen
  \bibfield  {author} {\bibinfo {author} {\bibfnamefont {A.}~\bibnamefont
  {Hosseinzadeh}}\ and\ \bibinfo {author} {\bibfnamefont {S.~A.}\ \bibnamefont
  {Jafari}},\ }\emph {Quantum Integrability of 1D Ionic Hubbard Model},\ \href
  {\doibase https://doi.org/10.1002/andp.201900601} {\bibfield  {journal}
  {\bibinfo  {journal} {Annalen der Physik}\ }\textbf {\bibinfo {volume}
  {532}},\ \bibinfo {pages} {1900601} (\bibinfo {year} {2020})}\BibitemShut
  {NoStop}%
\bibitem [{\citenamefont {Kollath}\ \emph {et~al.}(2010)\citenamefont
  {Kollath}, \citenamefont {Roux}, \citenamefont {Biroli},\ and\ \citenamefont
  {Läuchli}}]{KollathLaeuchli2010}%
  \BibitemOpen
  \bibfield  {author} {\bibinfo {author} {\bibfnamefont {C.}~\bibnamefont
  {Kollath}}, \bibinfo {author} {\bibfnamefont {G.}~\bibnamefont {Roux}},
  \bibinfo {author} {\bibfnamefont {G.}~\bibnamefont {Biroli}}, \ and\ \bibinfo
  {author} {\bibfnamefont {A.~M.}\ \bibnamefont {Läuchli}},\ }\emph
  {Statistical properties of the spectrum of the extended
  Bose{\textendash}Hubbard model},\ \href {\doibase
  10.1088/1742-5468/2010/08/p08011} {\bibfield  {journal} {\bibinfo  {journal}
  {Journal of Statistical Mechanics: Theory and Experiment}\ }\textbf {\bibinfo
  {volume} {2010}},\ \bibinfo {pages} {P08011} (\bibinfo {year}
  {2010})}\BibitemShut {NoStop}%
\bibitem [{\citenamefont {Prosen}(1999)}]{Prosen1999}%
  \BibitemOpen
  \bibfield  {author} {\bibinfo {author} {\bibfnamefont {T.}\
  \bibnamefont {Prosen}},\ }\emph {Ergodic properties of a generic
  nonintegrable quantum many-body system in the thermodynamic limit},\ \href
  {\doibase 10.1103/PhysRevE.60.3949} {\bibfield  {journal} {\bibinfo
  {journal} {Phys. Rev. E}\ }\textbf {\bibinfo {volume} {60}},\ \bibinfo
  {pages} {3949} (\bibinfo {year} {1999})}\BibitemShut {NoStop}%
\bibitem [{\citenamefont {Haque}\ and\ \citenamefont
  {McClarty}(2019)}]{Haque2019}%
  \BibitemOpen
  \bibfield  {author} {\bibinfo {author} {\bibfnamefont {M.}~\bibnamefont
  {Haque}}\ and\ \bibinfo {author} {\bibfnamefont {P.~A.}\ \bibnamefont
  {McClarty}},\ }\emph {Eigenstate thermalization scaling in Majorana clusters:
  From chaotic to integrable Sachdev-Ye-Kitaev models},\ \href {\doibase
  10.1103/PhysRevB.100.115122} {\bibfield  {journal} {\bibinfo  {journal}
  {Phys. Rev. B}\ }\textbf {\bibinfo {volume} {100}},\ \bibinfo {pages}
  {115122} (\bibinfo {year} {2019})}\BibitemShut {NoStop}%
\bibitem [{\citenamefont {Gebhard}(1997)}]{Gebhard1997}%
  \BibitemOpen
  \bibfield  {author} {\bibinfo {author} {\bibfnamefont {F.}~\bibnamefont
  {Gebhard}},\ }\href {\doibase 10.1007/3-540-14858-2} {\emph {\bibinfo {title}
  {The {M}ott Metal-Insulating Transition}}}\ (\bibinfo  {publisher}
  {Springer},\ \bibinfo {year} {1997})\BibitemShut {NoStop}%
\bibitem [{\citenamefont {Essler}\ \emph {et~al.}(2005)\citenamefont {Essler},
  \citenamefont {Frahm}, \citenamefont {G\"ohmann}, \citenamefont {Kl\"umper},\
  and\ \citenamefont {Korepin}}]{EsslerKorepin2005}%
  \BibitemOpen
  \bibfield  {author} {\bibinfo {author} {\bibfnamefont {F.}~\bibnamefont
  {Essler}}, \bibinfo {author} {\bibfnamefont {H.}~\bibnamefont {Frahm}},
  \bibinfo {author} {\bibfnamefont {F.}~\bibnamefont {G\"ohmann}}, \bibinfo
  {author} {\bibfnamefont {A.}~\bibnamefont {Kl\"umper}}, \ and\ \bibinfo
  {author} {\bibfnamefont {V.}~\bibnamefont {Korepin}},\ }\href {\doibase
  10.1017/CBO9780511534843} {\emph {\bibinfo {title} {The One-Dimensional
  {H}ubbard Model}}}\ (\bibinfo  {publisher} {Cambridge University Press},\
  \bibinfo {year} {2005})\BibitemShut {NoStop}%
\bibitem [{\citenamefont {Giamarchi}(2004)}]{Giamarchibook}%
  \BibitemOpen
  \bibfield  {author} {\bibinfo {author} {\bibfnamefont {T.}~\bibnamefont
  {Giamarchi}},\ }\href {\doibase 10.1093/acprof:oso/9780198525004.001.0001}
  {\emph {\bibinfo {title} {Quantum Physics in One Dimension}}}\ (\bibinfo
  {publisher} {Oxford University Press},\ \bibinfo {address} {Oxford},\
  \bibinfo {year} {2004})\BibitemShut {NoStop}%
\bibitem [{\citenamefont {{Bruus, H.}}\ and\ \citenamefont {{Angl\`es d'Auriac,
  J.-C.}}(1996)}]{Bruus1996}%
  \BibitemOpen
  \bibfield  {author} {\bibinfo {author} {\bibnamefont {{Bruus, H.}}}\ and\
  \bibinfo {author} {\bibnamefont {{Angl\`es d'Auriac, J.-C.}}},\ }\emph {The
  spectrum of the two-dimensional Hubbard model at low filling},\ \href
  {\doibase 10.1209/epl/i1996-00113-x} {\bibfield  {journal} {\bibinfo
  {journal} {Europhys. Lett.}\ }\textbf {\bibinfo {volume} {35}},\ \bibinfo
  {pages} {321} (\bibinfo {year} {1996})}\BibitemShut {NoStop}%
\bibitem [{\citenamefont {Pertot}\ \emph {et~al.}(2014)\citenamefont {Pertot},
  \citenamefont {Sheikhan}, \citenamefont {Cocchi}, \citenamefont {Miller},
  \citenamefont {Bohn}, \citenamefont {Koschorreck}, \citenamefont {K\"ohl},\
  and\ \citenamefont {Kollath}}]{PertotKollath2014}%
  \BibitemOpen
  \bibfield  {author} {\bibinfo {author} {\bibfnamefont {D.}~\bibnamefont
  {Pertot}}, \bibinfo {author} {\bibfnamefont {A.}~\bibnamefont {Sheikhan}},
  \bibinfo {author} {\bibfnamefont {E.}~\bibnamefont {Cocchi}}, \bibinfo
  {author} {\bibfnamefont {L.~A.}\ \bibnamefont {Miller}}, \bibinfo {author}
  {\bibfnamefont {J.~E.}\ \bibnamefont {Bohn}}, \bibinfo {author}
  {\bibfnamefont {M.}~\bibnamefont {Koschorreck}}, \bibinfo {author}
  {\bibfnamefont {M.}~\bibnamefont {K\"ohl}}, \ and\ \bibinfo {author}
  {\bibfnamefont {C.}~\bibnamefont {Kollath}},\ }\emph {Relaxation Dynamics of
  a Fermi Gas in an Optical Superlattice},\ \href {\doibase
  10.1103/PhysRevLett.113.170403} {\bibfield  {journal} {\bibinfo  {journal}
  {Phys. Rev. Lett.}\ }\textbf {\bibinfo {volume} {113}},\ \bibinfo {pages}
  {170403} (\bibinfo {year} {2014})}\BibitemShut {NoStop}%
\bibitem [{\citenamefont {Messer}\ \emph {et~al.}(2015)\citenamefont {Messer},
  \citenamefont {Desbuquois}, \citenamefont {Uehlinger}, \citenamefont {Jotzu},
  \citenamefont {Huber}, \citenamefont {Greif},\ and\ \citenamefont
  {Esslinger}}]{MesserEsslinger2015}%
  \BibitemOpen
  \bibfield  {author} {\bibinfo {author} {\bibfnamefont {M.}~\bibnamefont
  {Messer}}, \bibinfo {author} {\bibfnamefont {R.}~\bibnamefont {Desbuquois}},
  \bibinfo {author} {\bibfnamefont {T.}~\bibnamefont {Uehlinger}}, \bibinfo
  {author} {\bibfnamefont {G.}~\bibnamefont {Jotzu}}, \bibinfo {author}
  {\bibfnamefont {S.}~\bibnamefont {Huber}}, \bibinfo {author} {\bibfnamefont
  {D.}~\bibnamefont {Greif}}, \ and\ \bibinfo {author} {\bibfnamefont
  {T.}~\bibnamefont {Esslinger}},\ }\emph {Exploring Competing Density Order in
  the Ionic Hubbard Model with Ultracold Fermions},\ \href {\doibase
  10.1103/PhysRevLett.115.115303} {\bibfield  {journal} {\bibinfo  {journal}
  {Phys. Rev. Lett.}\ }\textbf {\bibinfo {volume} {115}},\ \bibinfo {pages}
  {115303} (\bibinfo {year} {2015})}\BibitemShut {NoStop}%
\bibitem [{\citenamefont {Kennes}\ \emph {et~al.}(2020)\citenamefont {Kennes},
  \citenamefont {Xian}, \citenamefont {Claassen},\ and\ \citenamefont
  {Rubio}}]{KennesRubio2020}%
  \BibitemOpen
  \bibfield  {author} {\bibinfo {author} {\bibfnamefont {D.~M.}\ \bibnamefont
  {Kennes}}, \bibinfo {author} {\bibfnamefont {L.}~\bibnamefont {Xian}},
  \bibinfo {author} {\bibfnamefont {M.}~\bibnamefont {Claassen}}, \ and\
  \bibinfo {author} {\bibfnamefont {A.}~\bibnamefont {Rubio}},\ }\emph
  {One-dimensional flat bands in twisted bilayer germanium selenide},\ \href
  {\doibase 10.1038/s41467-020-14947-0} {\bibfield  {journal} {\bibinfo
  {journal} {Nature Communications}\ }\textbf {\bibinfo {volume} {11}},\
  \bibinfo {pages} {1124} (\bibinfo {year} {2020})}\BibitemShut {NoStop}%
\bibitem [{\citenamefont {Torrance}\ \emph {et~al.}(1981)\citenamefont
  {Torrance}, \citenamefont {Girlando}, \citenamefont {Mayerle}, \citenamefont
  {Crowley}, \citenamefont {Lee}, \citenamefont {Batail},\ and\ \citenamefont
  {LaPlaca}}]{TorranceLaPlaca1981}%
  \BibitemOpen
  \bibfield  {author} {\bibinfo {author} {\bibfnamefont {J.~B.}\ \bibnamefont
  {Torrance}}, \bibinfo {author} {\bibfnamefont {A.}~\bibnamefont {Girlando}},
  \bibinfo {author} {\bibfnamefont {J.~J.}\ \bibnamefont {Mayerle}}, \bibinfo
  {author} {\bibfnamefont {J.~I.}\ \bibnamefont {Crowley}}, \bibinfo {author}
  {\bibfnamefont {V.~Y.}\ \bibnamefont {Lee}}, \bibinfo {author} {\bibfnamefont
  {P.}~\bibnamefont {Batail}}, \ and\ \bibinfo {author} {\bibfnamefont {S.~J.}\
  \bibnamefont {LaPlaca}},\ }\emph {Anomalous Nature of Neutral-to-Ionic Phase
  Transition in Tetrathiafulvalene-Chloranil},\ \href {\doibase
  10.1103/PhysRevLett.47.1747} {\bibfield  {journal} {\bibinfo  {journal}
  {Phys. Rev. Lett.}\ }\textbf {\bibinfo {volume} {47}},\ \bibinfo {pages}
  {1747} (\bibinfo {year} {1981})}\BibitemShut {NoStop}%
\bibitem [{\citenamefont {Nagaosa}\ and\ \citenamefont
  {Takimoto}(1986)}]{NagaosaTakimoto1986}%
  \BibitemOpen
  \bibfield  {author} {\bibinfo {author} {\bibfnamefont {N.}~\bibnamefont
  {Nagaosa}}\ and\ \bibinfo {author} {\bibfnamefont {J.-i.}\ \bibnamefont
  {Takimoto}},\ }\emph {Theory of Neutral-Ionic Transition in Organic Crystals.
  I. Monte Carlo Simulation of Modified Hubbard Model},\ \href {\doibase
  10.1143/JPSJ.55.2735} {\bibfield  {journal} {\bibinfo  {journal} {Journal of
  the Physical Society of Japan}\ }\textbf {\bibinfo {volume} {55}},\ \bibinfo
  {pages} {2735} (\bibinfo {year} {1986})}\BibitemShut {NoStop}%
\bibitem [{\citenamefont {Egami}\ \emph {et~al.}(1993)\citenamefont {Egami},
  \citenamefont {Ishihara},\ and\ \citenamefont {Tachiki}}]{EgamiTachiki1993}%
  \BibitemOpen
  \bibfield  {author} {\bibinfo {author} {\bibfnamefont {T.}~\bibnamefont
  {Egami}}, \bibinfo {author} {\bibfnamefont {S.}~\bibnamefont {Ishihara}}, \
  and\ \bibinfo {author} {\bibfnamefont {M.}~\bibnamefont {Tachiki}},\ }\emph
  {Lattice Effect of Strong Electron Correlation: Implication for
  Ferroelectricity and Superconductivity},\ \href {\doibase
  10.1126/science.261.5126.1307} {\bibfield  {journal} {\bibinfo  {journal}
  {Science}\ }\textbf {\bibinfo {volume} {261}},\ \bibinfo {pages} {1307}
  (\bibinfo {year} {1993})}\BibitemShut {NoStop}%
\bibitem [{\citenamefont {Kogut}\ and\ \citenamefont
  {Susskind}(1975)}]{Kogut1975}%
  \BibitemOpen
  \bibfield  {author} {\bibinfo {author} {\bibfnamefont {J.}~\bibnamefont
  {Kogut}}\ and\ \bibinfo {author} {\bibfnamefont {L.}~\bibnamefont
  {Susskind}},\ }\emph {Hamiltonian formulation of Wilson's lattice gauge
  theories},\ \href {\doibase 10.1103/PhysRevD.11.395} {\bibfield  {journal}
  {\bibinfo  {journal} {Phys. Rev. D}\ }\textbf {\bibinfo {volume} {11}},\
  \bibinfo {pages} {395} (\bibinfo {year} {1975})}\BibitemShut {NoStop}%
\bibitem [{\citenamefont {Fabrizio}\ \emph {et~al.}(1999)\citenamefont
  {Fabrizio}, \citenamefont {Gogolin},\ and\ \citenamefont
  {Nersesyan}}]{FabrizioNersesyan1999}%
  \BibitemOpen
  \bibfield  {author} {\bibinfo {author} {\bibfnamefont {M.}~\bibnamefont
  {Fabrizio}}, \bibinfo {author} {\bibfnamefont {A.~O.}\ \bibnamefont
  {Gogolin}}, \ and\ \bibinfo {author} {\bibfnamefont {A.~A.}\ \bibnamefont
  {Nersesyan}},\ }\emph {From Band Insulator to Mott Insulator in One
  Dimension},\ \href {\doibase 10.1103/PhysRevLett.83.2014} {\bibfield
  {journal} {\bibinfo  {journal} {Phys. Rev. Lett.}\ }\textbf {\bibinfo
  {volume} {83}},\ \bibinfo {pages} {2014} (\bibinfo {year}
  {1999})}\BibitemShut {NoStop}%
\bibitem [{\citenamefont {Fabrizio}\ \emph {et~al.}(2000)\citenamefont
  {Fabrizio}, \citenamefont {Gogolin},\ and\ \citenamefont
  {Nersesyan}}]{FabrizioNersesyan2000}%
  \BibitemOpen
  \bibfield  {author} {\bibinfo {author} {\bibfnamefont {M.}~\bibnamefont
  {Fabrizio}}, \bibinfo {author} {\bibfnamefont {A.}~\bibnamefont {Gogolin}}, \
  and\ \bibinfo {author} {\bibfnamefont {A.}~\bibnamefont {Nersesyan}},\ }\emph
  {Critical properties of the double-frequency sine-Gordon model with
  applications},\ \href {\doibase
  https://doi.org/10.1016/S0550-3213(00)00247-9} {\bibfield  {journal}
  {\bibinfo  {journal} {Nuclear Physics B}\ }\textbf {\bibinfo {volume}
  {580}},\ \bibinfo {pages} {647} (\bibinfo {year} {2000})}\BibitemShut
  {NoStop}%
\bibitem [{\citenamefont {Gidopoulos}\ \emph {et~al.}(2000)\citenamefont
  {Gidopoulos}, \citenamefont {Sorella},\ and\ \citenamefont
  {Tosatti}}]{GidopoulosTosatti2000}%
  \BibitemOpen
  \bibfield  {author} {\bibinfo {author} {\bibfnamefont {N.}~\bibnamefont
  {Gidopoulos}}, \bibinfo {author} {\bibfnamefont {S.}~\bibnamefont {Sorella}},
  \ and\ \bibinfo {author} {\bibfnamefont {E.}~\bibnamefont {Tosatti}},\ }\emph
  {Born effective charge reversal and metallic threshold state at a band
  insulator-Mott insulator transition},\ \href {\doibase 10.1007/s100510050123}
  {\bibfield  {journal} {\bibinfo  {journal} {The European Physical Journal B -
  Condensed Matter and Complex Systems}\ }\textbf {\bibinfo {volume} {14}},\
  \bibinfo {pages} {217} (\bibinfo {year} {2000})}\BibitemShut {NoStop}%
\bibitem [{\citenamefont {Torio}\ \emph {et~al.}(2001)\citenamefont {Torio},
  \citenamefont {Aligia},\ and\ \citenamefont {Ceccatto}}]{TorioCeccatto2001}%
  \BibitemOpen
  \bibfield  {author} {\bibinfo {author} {\bibfnamefont {M.~E.}\ \bibnamefont
  {Torio}}, \bibinfo {author} {\bibfnamefont {A.~A.}\ \bibnamefont {Aligia}}, \
  and\ \bibinfo {author} {\bibfnamefont {H.~A.}\ \bibnamefont {Ceccatto}},\
  }\emph {Phase diagram of the Hubbard chain with two atoms per cell},\ \href
  {\doibase 10.1103/PhysRevB.64.121105} {\bibfield  {journal} {\bibinfo
  {journal} {Phys. Rev. B}\ }\textbf {\bibinfo {volume} {64}},\ \bibinfo
  {pages} {121105} (\bibinfo {year} {2001})}\BibitemShut {NoStop}%
\bibitem [{\citenamefont {Wilkens}\ and\ \citenamefont
  {Martin}(2001)}]{WilkensMartin2001}%
  \BibitemOpen
  \bibfield  {author} {\bibinfo {author} {\bibfnamefont {T.}~\bibnamefont
  {Wilkens}}\ and\ \bibinfo {author} {\bibfnamefont {R.~M.}\ \bibnamefont
  {Martin}},\ }\emph {Quantum Monte Carlo study of the one-dimensional ionic
  Hubbard model},\ \href {\doibase 10.1103/PhysRevB.63.235108} {\bibfield
  {journal} {\bibinfo  {journal} {Phys. Rev. B}\ }\textbf {\bibinfo {volume}
  {63}},\ \bibinfo {pages} {235108} (\bibinfo {year} {2001})}\BibitemShut
  {NoStop}%
\bibitem [{\citenamefont {Kampf}\ \emph {et~al.}(2003)\citenamefont {Kampf},
  \citenamefont {Sekania}, \citenamefont {Japaridze},\ and\ \citenamefont
  {Brune}}]{KampfBrune2003}%
  \BibitemOpen
  \bibfield  {author} {\bibinfo {author} {\bibfnamefont {A.~P.}\ \bibnamefont
  {Kampf}}, \bibinfo {author} {\bibfnamefont {M.}~\bibnamefont {Sekania}},
  \bibinfo {author} {\bibfnamefont {G.~I.}\ \bibnamefont {Japaridze}}, \ and\
  \bibinfo {author} {\bibfnamefont {P.}~\bibnamefont {Brune}},\ }\emph {Nature
  of the insulating phases in the half-filled ionic Hubbard model},\ \href
  {\doibase 10.1088/0953-8984/15/34/319} {\bibfield  {journal} {\bibinfo
  {journal} {Journal of Physics: Condensed Matter}\ }\textbf {\bibinfo {volume}
  {15}},\ \bibinfo {pages} {5895} (\bibinfo {year} {2003})}\BibitemShut
  {NoStop}%
\bibitem [{\citenamefont {Zhang}\ \emph {et~al.}(2003)\citenamefont {Zhang},
  \citenamefont {Wu},\ and\ \citenamefont {Lin}}]{ZhangLin2003}%
  \BibitemOpen
  \bibfield  {author} {\bibinfo {author} {\bibfnamefont {Y.~Z.}\ \bibnamefont
  {Zhang}}, \bibinfo {author} {\bibfnamefont {C.~Q.}\ \bibnamefont {Wu}}, \
  and\ \bibinfo {author} {\bibfnamefont {H.~Q.}\ \bibnamefont {Lin}},\ }\emph
  {Inducement of bond-order wave due to electron correlation in one
  dimension},\ \href {\doibase 10.1103/PhysRevB.67.205109} {\bibfield
  {journal} {\bibinfo  {journal} {Phys. Rev. B}\ }\textbf {\bibinfo {volume}
  {67}},\ \bibinfo {pages} {205109} (\bibinfo {year} {2003})}\BibitemShut
  {NoStop}%
\bibitem [{\citenamefont {Manmana}\ \emph {et~al.}(2004)\citenamefont
  {Manmana}, \citenamefont {Meden}, \citenamefont {Noack},\ and\ \citenamefont
  {Sch\"onhammer}}]{ManmanaSchonhammer2004}%
  \BibitemOpen
  \bibfield  {author} {\bibinfo {author} {\bibfnamefont {S.~R.}\ \bibnamefont
  {Manmana}}, \bibinfo {author} {\bibfnamefont {V.}~\bibnamefont {Meden}},
  \bibinfo {author} {\bibfnamefont {R.~M.}\ \bibnamefont {Noack}}, \ and\
  \bibinfo {author} {\bibfnamefont {K.}~\bibnamefont {Sch\"onhammer}},\ }\emph
  {Quantum critical behavior of the one-dimensional ionic Hubbard model},\
  \href {\doibase 10.1103/PhysRevB.70.155115} {\bibfield  {journal} {\bibinfo
  {journal} {Phys. Rev. B}\ }\textbf {\bibinfo {volume} {70}},\ \bibinfo
  {pages} {155115} (\bibinfo {year} {2004})}\BibitemShut {NoStop}%
\bibitem [{\citenamefont {Batista}\ and\ \citenamefont
  {Aligia}(2004)}]{BatistaAligia2004}%
  \BibitemOpen
  \bibfield  {author} {\bibinfo {author} {\bibfnamefont {C.~D.}\ \bibnamefont
  {Batista}}\ and\ \bibinfo {author} {\bibfnamefont {A.~A.}\ \bibnamefont
  {Aligia}},\ }\emph {Exact Bond Ordered Ground State for the Transition
  between the Band and the Mott Insulator},\ \href {\doibase
  10.1103/PhysRevLett.92.246405} {\bibfield  {journal} {\bibinfo  {journal}
  {Phys. Rev. Lett.}\ }\textbf {\bibinfo {volume} {92}},\ \bibinfo {pages}
  {246405} (\bibinfo {year} {2004})}\BibitemShut {NoStop}%
\bibitem [{\citenamefont {Otsuka}\ and\ \citenamefont
  {Nakamura}(2005)}]{OtsukaNakamura2005}%
  \BibitemOpen
  \bibfield  {author} {\bibinfo {author} {\bibfnamefont {H.}~\bibnamefont
  {Otsuka}}\ and\ \bibinfo {author} {\bibfnamefont {M.}~\bibnamefont
  {Nakamura}},\ }\emph {Ground-state phase diagram of the one-dimensional
  Hubbard model with an alternating chemical potential},\ \href {\doibase
  10.1103/PhysRevB.71.155105} {\bibfield  {journal} {\bibinfo  {journal} {Phys.
  Rev. B}\ }\textbf {\bibinfo {volume} {71}},\ \bibinfo {pages} {155105}
  (\bibinfo {year} {2005})}\BibitemShut {NoStop}%
\bibitem [{\citenamefont {Refolio}\ \emph {et~al.}(2005)\citenamefont
  {Refolio}, \citenamefont {Sancho},\ and\ \citenamefont
  {Rubio}}]{RefolioRubio2005}%
  \BibitemOpen
  \bibfield  {author} {\bibinfo {author} {\bibfnamefont {M.~C.}\ \bibnamefont
  {Refolio}}, \bibinfo {author} {\bibfnamefont {J.~M.~L.}\ \bibnamefont
  {Sancho}}, \ and\ \bibinfo {author} {\bibfnamefont {J.}~\bibnamefont
  {Rubio}},\ }\emph {Modelling one-dimensional insulating materials with the
  ionic Hubbard model},\ \href {\doibase 10.1088/0953-8984/17/42/004}
  {\bibfield  {journal} {\bibinfo  {journal} {Journal of Physics: Condensed
  Matter}\ }\textbf {\bibinfo {volume} {17}},\ \bibinfo {pages} {6635}
  (\bibinfo {year} {2005})}\BibitemShut {NoStop}%
\bibitem [{\citenamefont {Aligia}\ and\ \citenamefont
  {Batista}(2005)}]{AligiaBatista2005}%
  \BibitemOpen
  \bibfield  {author} {\bibinfo {author} {\bibfnamefont {A.~A.}\ \bibnamefont
  {Aligia}}\ and\ \bibinfo {author} {\bibfnamefont {C.~D.}\ \bibnamefont
  {Batista}},\ }\emph {Dimerized phase of ionic Hubbard models},\ \href
  {\doibase 10.1103/PhysRevB.71.125110} {\bibfield  {journal} {\bibinfo
  {journal} {Phys. Rev. B}\ }\textbf {\bibinfo {volume} {71}},\ \bibinfo
  {pages} {125110} (\bibinfo {year} {2005})}\BibitemShut {NoStop}%
\bibitem [{\citenamefont {Legeza}\ \emph {et~al.}(2006)\citenamefont {Legeza},
  \citenamefont {Buchta},\ and\ \citenamefont {S\'olyom}}]{LegezaSolyom2006}%
  \BibitemOpen
  \bibfield  {author} {\bibinfo {author} {\bibfnamefont {O.}~\bibnamefont
  {Legeza}}, \bibinfo {author} {\bibfnamefont {K.}~\bibnamefont {Buchta}}, \
  and\ \bibinfo {author} {\bibfnamefont {J.}~\bibnamefont {S\'olyom}},\ }\emph
  {Unified phase diagram of models exhibiting a neutral-ionic transition},\
  \href {\doibase 10.1103/PhysRevB.73.165124} {\bibfield  {journal} {\bibinfo
  {journal} {Phys. Rev. B}\ }\textbf {\bibinfo {volume} {73}},\ \bibinfo
  {pages} {165124} (\bibinfo {year} {2006})}\BibitemShut {NoStop}%
\bibitem [{\citenamefont {Tincani}\ \emph {et~al.}(2009)\citenamefont
  {Tincani}, \citenamefont {Noack},\ and\ \citenamefont
  {Baeriswyl}}]{TincaniBaeriswyl2009}%
  \BibitemOpen
  \bibfield  {author} {\bibinfo {author} {\bibfnamefont {L.}~\bibnamefont
  {Tincani}}, \bibinfo {author} {\bibfnamefont {R.~M.}\ \bibnamefont {Noack}},
  \ and\ \bibinfo {author} {\bibfnamefont {D.}~\bibnamefont {Baeriswyl}},\
  }\emph {Critical properties of the band-insulator-to-Mott-insulator
  transition in the strong-coupling limit of the ionic Hubbard model},\ \href
  {\doibase 10.1103/PhysRevB.79.165109} {\bibfield  {journal} {\bibinfo
  {journal} {Phys. Rev. B}\ }\textbf {\bibinfo {volume} {79}},\ \bibinfo
  {pages} {165109} (\bibinfo {year} {2009})}\BibitemShut {NoStop}%
\bibitem [{\citenamefont {Hafez}\ and\ \citenamefont
  {Jafari}(2010)}]{HafezJafari2010}%
  \BibitemOpen
  \bibfield  {author} {\bibinfo {author} {\bibfnamefont {M.}~\bibnamefont
  {Hafez}}\ and\ \bibinfo {author} {\bibfnamefont {S.~A.}\ \bibnamefont
  {Jafari}},\ }\emph {Excitation spectrum of one-dimensional extended ionic
  Hubbard model},\ \href {\doibase 10.1140/epjb/e2010-10509-x} {\bibfield
  {journal} {\bibinfo  {journal} {The European Physical Journal B}\ }\textbf
  {\bibinfo {volume} {78}},\ \bibinfo {pages} {323} (\bibinfo {year}
  {2010})}\BibitemShut {NoStop}%
\bibitem [{\citenamefont {Go}\ and\ \citenamefont {Jeon}(2011)}]{GoJeon2011}%
  \BibitemOpen
  \bibfield  {author} {\bibinfo {author} {\bibfnamefont {A.}~\bibnamefont
  {Go}}\ and\ \bibinfo {author} {\bibfnamefont {G.~S.}\ \bibnamefont {Jeon}},\
  }\emph {Phase transitions and spectral properties of the ionic Hubbard model
  in one dimension},\ \href {\doibase 10.1103/PhysRevB.84.195102} {\bibfield
  {journal} {\bibinfo  {journal} {Phys. Rev. B}\ }\textbf {\bibinfo {volume}
  {84}},\ \bibinfo {pages} {195102} (\bibinfo {year} {2011})}\BibitemShut
  {NoStop}%
\bibitem [{\citenamefont {Kim}\ \emph {et~al.}(2014)\citenamefont {Kim},
  \citenamefont {Choi},\ and\ \citenamefont {Jeon}}]{KimJeon2014}%
  \BibitemOpen
  \bibfield  {author} {\bibinfo {author} {\bibfnamefont {A.~J.}\ \bibnamefont
  {Kim}}, \bibinfo {author} {\bibfnamefont {M.~Y.}\ \bibnamefont {Choi}}, \
  and\ \bibinfo {author} {\bibfnamefont {G.~S.}\ \bibnamefont {Jeon}},\ }\emph
  {Finite-temperature phase transitions in the ionic Hubbard model},\ \href
  {\doibase 10.1103/PhysRevB.89.165117} {\bibfield  {journal} {\bibinfo
  {journal} {Phys. Rev. B}\ }\textbf {\bibinfo {volume} {89}},\ \bibinfo
  {pages} {165117} (\bibinfo {year} {2014})}\BibitemShut {NoStop}%
\bibitem [{\citenamefont {Hafez~Torbati}\ \emph {et~al.}(2014)\citenamefont
  {Hafez~Torbati}, \citenamefont {Drescher},\ and\ \citenamefont
  {Uhrig}}]{HafezTorbatiUhrig2014}%
  \BibitemOpen
  \bibfield  {author} {\bibinfo {author} {\bibfnamefont {M.}~\bibnamefont
  {Hafez~Torbati}}, \bibinfo {author} {\bibfnamefont {N.~A.}\ \bibnamefont
  {Drescher}}, \ and\ \bibinfo {author} {\bibfnamefont {G.~S.}\ \bibnamefont
  {Uhrig}},\ }\emph {Dispersive excitations in one-dimensional ionic Hubbard
  model},\ \href {\doibase 10.1103/PhysRevB.89.245126} {\bibfield  {journal}
  {\bibinfo  {journal} {Phys. Rev. B}\ }\textbf {\bibinfo {volume} {89}},\
  \bibinfo {pages} {245126} (\bibinfo {year} {2014})}\BibitemShut {NoStop}%
\bibitem [{\citenamefont {Hafez-Torbati}\ \emph {et~al.}(2015)\citenamefont
  {Hafez-Torbati}, \citenamefont {Drescher},\ and\ \citenamefont
  {Uhrig}}]{HafezTorbatiUhrig2015}%
  \BibitemOpen
  \bibfield  {author} {\bibinfo {author} {\bibfnamefont {M.}~\bibnamefont
  {Hafez-Torbati}}, \bibinfo {author} {\bibfnamefont {N.~A.}\ \bibnamefont
  {Drescher}}, \ and\ \bibinfo {author} {\bibfnamefont {G.~S.}\ \bibnamefont
  {Uhrig}},\ }\emph {From gapped excitons to gapless triplons in one
  dimension},\ \href {\doibase 10.1140/epjb/e2014-50551-0} {\bibfield
  {journal} {\bibinfo  {journal} {The European Physical Journal B}\ }\textbf
  {\bibinfo {volume} {88}},\ \bibinfo {pages} {3} (\bibinfo {year}
  {2015})}\BibitemShut {NoStop}%
\bibitem [{\citenamefont {Bag}\ \emph {et~al.}(2015)\citenamefont {Bag},
  \citenamefont {Garg},\ and\ \citenamefont
  {Krishnamurthy}}]{BagKrishnamurthy2015}%
  \BibitemOpen
  \bibfield  {author} {\bibinfo {author} {\bibfnamefont {S.}~\bibnamefont
  {Bag}}, \bibinfo {author} {\bibfnamefont {A.}~\bibnamefont {Garg}}, \ and\
  \bibinfo {author} {\bibfnamefont {H.~R.}\ \bibnamefont {Krishnamurthy}},\
  }\emph {Phase diagram of the half-filled ionic Hubbard model},\ \href
  {\doibase 10.1103/PhysRevB.91.235108} {\bibfield  {journal} {\bibinfo
  {journal} {Phys. Rev. B}\ }\textbf {\bibinfo {volume} {91}},\ \bibinfo
  {pages} {235108} (\bibinfo {year} {2015})}\BibitemShut {NoStop}%
\bibitem [{\citenamefont {Loida}\ \emph {et~al.}(2017)\citenamefont {Loida},
  \citenamefont {Bernier}, \citenamefont {Citro}, \citenamefont {Orignac},\
  and\ \citenamefont {Kollath}}]{LoidaKollath2017}%
  \BibitemOpen
  \bibfield  {author} {\bibinfo {author} {\bibfnamefont {K.}~\bibnamefont
  {Loida}}, \bibinfo {author} {\bibfnamefont {J.-S.}\ \bibnamefont {Bernier}},
  \bibinfo {author} {\bibfnamefont {R.}~\bibnamefont {Citro}}, \bibinfo
  {author} {\bibfnamefont {E.}~\bibnamefont {Orignac}}, \ and\ \bibinfo
  {author} {\bibfnamefont {C.}~\bibnamefont {Kollath}},\ }\emph {Probing the
  Bond Order Wave Phase Transitions of the Ionic Hubbard Model by Superlattice
  Modulation Spectroscopy},\ \href {\doibase 10.1103/PhysRevLett.119.230403}
  {\bibfield  {journal} {\bibinfo  {journal} {Phys. Rev. Lett.}\ }\textbf
  {\bibinfo {volume} {119}},\ \bibinfo {pages} {230403} (\bibinfo {year}
  {2017})}\BibitemShut {NoStop}%
\bibitem [{\citenamefont {Chattopadhyay}\ \emph {et~al.}(2019)\citenamefont
  {Chattopadhyay}, \citenamefont {Bag}, \citenamefont {Krishnamurthy},\ and\
  \citenamefont {Garg}}]{ChattopadhyayGarg2019}%
  \BibitemOpen
  \bibfield  {author} {\bibinfo {author} {\bibfnamefont {A.}~\bibnamefont
  {Chattopadhyay}}, \bibinfo {author} {\bibfnamefont {S.}~\bibnamefont {Bag}},
  \bibinfo {author} {\bibfnamefont {H.~R.}\ \bibnamefont {Krishnamurthy}}, \
  and\ \bibinfo {author} {\bibfnamefont {A.}~\bibnamefont {Garg}},\ }\emph
  {Phase diagram of the half-filled ionic Hubbard model in the limit of strong
  correlations},\ \href {\doibase 10.1103/PhysRevB.99.155127} {\bibfield
  {journal} {\bibinfo  {journal} {Phys. Rev. B}\ }\textbf {\bibinfo {volume}
  {99}},\ \bibinfo {pages} {155127} (\bibinfo {year} {2019})}\BibitemShut
  {NoStop}%
\bibitem [{\citenamefont {Bohigas}\ \emph {et~al.}(1986)\citenamefont
  {Bohigas}, \citenamefont {Giannoni},\ and\ \citenamefont
  {Schmit}}]{BohigasGiannoni1986}%
  \BibitemOpen
  \bibfield  {author} {\bibinfo {author} {\bibfnamefont {O.}~\bibnamefont
  {Bohigas}}, \bibinfo {author} {\bibfnamefont {M.-J.}\ \bibnamefont
  {Giannoni}}, \ and\ \bibinfo {author} {\bibfnamefont {C.}~\bibnamefont
  {Schmit}},\ }\href {\doibase 10.1007/3-540-17171-1_2} {\emph {\bibinfo
  {title} {Quantum Chaos and Statistical Nuclear Physics}}},\ edited by\
  \bibinfo {editor} {\bibfnamefont {T.~H.}\ \bibnamefont {Seligman}}\ and\
  \bibinfo {editor} {\bibfnamefont {H.}~\bibnamefont {Nishioka}}\ (\bibinfo
  {publisher} {Springer Berlin Heidelberg},\ \bibinfo {address} {Berlin,
  Heidelberg},\ \bibinfo {year} {1986})\ pp.\ \bibinfo {pages}
  {18--40}\BibitemShut {NoStop}%
\bibitem [{\citenamefont {Guhr}\ \emph {et~al.}(1998)\citenamefont {Guhr},
  \citenamefont {M\"uller-Groeling},\ and\ \citenamefont
  {Weidenm\"uller}}]{GuhrWeidenmueller1998}%
  \BibitemOpen
  \bibfield  {author} {\bibinfo {author} {\bibfnamefont {T.}~\bibnamefont
  {Guhr}}, \bibinfo {author} {\bibfnamefont {A.}~\bibnamefont
  {M\"uller-Groeling}}, \ and\ \bibinfo {author} {\bibfnamefont {H.~A.}\
  \bibnamefont {Weidenm\"uller}},\ }\emph {Random-matrix theories in quantum
  physics: common concepts},\ \href {\doibase
  https://doi.org/10.1016/S0370-1573(97)00088-4} {\bibfield  {journal}
  {\bibinfo  {journal} {Physics Reports}\ }\textbf {\bibinfo {volume} {299}},\
  \bibinfo {pages} {189} (\bibinfo {year} {1998})}\BibitemShut {NoStop}%
\bibitem [{\citenamefont {Hsu}\ and\ \citenamefont
  {Angle`s~d'Auriac}(1993)}]{HsuAngle1993}%
  \BibitemOpen
  \bibfield  {author} {\bibinfo {author} {\bibfnamefont {T.~C.}\ \bibnamefont
  {Hsu}}\ and\ \bibinfo {author} {\bibfnamefont {J.~C.}\ \bibnamefont
  {Angle`s~d'Auriac}},\ }\emph {Level repulsion in integrable and
  almost-integrable quantum spin models},\ \href {\doibase
  10.1103/PhysRevB.47.14291} {\bibfield  {journal} {\bibinfo  {journal} {Phys.
  Rev. B}\ }\textbf {\bibinfo {volume} {47}},\ \bibinfo {pages} {14291}
  (\bibinfo {year} {1993})}\BibitemShut {NoStop}%
\bibitem [{\citenamefont {Montambaux}\ \emph {et~al.}(1993)\citenamefont
  {Montambaux}, \citenamefont {Poilblanc}, \citenamefont {Bellissard},\ and\
  \citenamefont {Sire}}]{MontambauxSire1993}%
  \BibitemOpen
  \bibfield  {author} {\bibinfo {author} {\bibfnamefont {G.}~\bibnamefont
  {Montambaux}}, \bibinfo {author} {\bibfnamefont {D.}~\bibnamefont
  {Poilblanc}}, \bibinfo {author} {\bibfnamefont {J.}~\bibnamefont
  {Bellissard}}, \ and\ \bibinfo {author} {\bibfnamefont {C.}~\bibnamefont
  {Sire}},\ }\emph {Quantum chaos in spin-fermion models},\ \href {\doibase
  10.1103/PhysRevLett.70.497} {\bibfield  {journal} {\bibinfo  {journal} {Phys.
  Rev. Lett.}\ }\textbf {\bibinfo {volume} {70}},\ \bibinfo {pages} {497}
  (\bibinfo {year} {1993})}\BibitemShut {NoStop}%
\bibitem [{\citenamefont {Oganesyan}\ and\ \citenamefont
  {Huse}(2007)}]{OganesyanHuse2007}%
  \BibitemOpen
  \bibfield  {author} {\bibinfo {author} {\bibfnamefont {V.}~\bibnamefont
  {Oganesyan}}\ and\ \bibinfo {author} {\bibfnamefont {D.~A.}\ \bibnamefont
  {Huse}},\ }\emph {Localization of interacting fermions at high temperature},\
  \href {http://link.aps.org/abstract/PRB/v75/e155111} {\bibfield  {journal}
  {\bibinfo  {journal} {Physical Review B (Condensed Matter and Materials
  Physics)}\ }\textbf {\bibinfo {volume} {75}},\ \bibinfo {eid} {155111}
  (\bibinfo {year} {2007})}\BibitemShut {NoStop}%
\bibitem [{\citenamefont {Serbyn}\ and\ \citenamefont
  {Moore}(2016)}]{SerbynMoore2016}%
  \BibitemOpen
  \bibfield  {author} {\bibinfo {author} {\bibfnamefont {M.}~\bibnamefont
  {Serbyn}}\ and\ \bibinfo {author} {\bibfnamefont {J.~E.}\ \bibnamefont
  {Moore}},\ }\emph {Spectral statistics across the many-body localization
  transition},\ \href {\doibase 10.1103/PhysRevB.93.041424} {\bibfield
  {journal} {\bibinfo  {journal} {Phys. Rev. B}\ }\textbf {\bibinfo {volume}
  {93}},\ \bibinfo {pages} {041424} (\bibinfo {year} {2016})}\BibitemShut
  {NoStop}%
\bibitem [{\citenamefont {Vyas}\ and\ \citenamefont
  {Seligman}(2018)}]{MananSeligman2018}%
  \BibitemOpen
  \bibfield  {author} {\bibinfo {author} {\bibfnamefont {M.}~\bibnamefont
  {Vyas}}\ and\ \bibinfo {author} {\bibfnamefont {T.~H.}\ \bibnamefont
  {Seligman}},\ }\emph {Random matrix ensembles for many-body quantum
  systems},\ \href {\doibase 10.1063/1.5031701} {\bibfield  {journal} {\bibinfo
   {journal} {AIP Conference Proceedings}\ }\textbf {\bibinfo {volume}
  {1950}},\ \bibinfo {pages} {030009} (\bibinfo {year} {2018})},\ \Eprint
  {http://arxiv.org/abs/https://aip.scitation.org/doi/pdf/10.1063/1.5031701}
  {https://aip.scitation.org/doi/pdf/10.1063/1.5031701} \BibitemShut {NoStop}%
\bibitem [{\citenamefont {G\'omez}\ \emph {et~al.}(2002)\citenamefont
  {G\'omez}, \citenamefont {Molina}, \citenamefont {Rela\~no},\ and\
  \citenamefont {Retamosa}}]{GomezRetamosa2002}%
  \BibitemOpen
  \bibfield  {author} {\bibinfo {author} {\bibfnamefont {J.~M.~G.}\
  \bibnamefont {G\'omez}}, \bibinfo {author} {\bibfnamefont {R.~A.}\
  \bibnamefont {Molina}}, \bibinfo {author} {\bibfnamefont {A.}~\bibnamefont
  {Rela\~no}}, \ and\ \bibinfo {author} {\bibfnamefont {J.}~\bibnamefont
  {Retamosa}},\ }\emph {Misleading signatures of quantum chaos},\ \href
  {\doibase 10.1103/PhysRevE.66.036209} {\bibfield  {journal} {\bibinfo
  {journal} {Phys. Rev. E}\ }\textbf {\bibinfo {volume} {66}},\ \bibinfo
  {pages} {036209} (\bibinfo {year} {2002})}\BibitemShut {NoStop}%
\bibitem [{\citenamefont {Atas}\ \emph {et~al.}(2013)\citenamefont {Atas},
  \citenamefont {Bogomolny}, \citenamefont {Giraud},\ and\ \citenamefont
  {Roux}}]{AtasRoux2013}%
  \BibitemOpen
  \bibfield  {author} {\bibinfo {author} {\bibfnamefont {Y.~Y.}\ \bibnamefont
  {Atas}}, \bibinfo {author} {\bibfnamefont {E.}~\bibnamefont {Bogomolny}},
  \bibinfo {author} {\bibfnamefont {O.}~\bibnamefont {Giraud}}, \ and\ \bibinfo
  {author} {\bibfnamefont {G.}~\bibnamefont {Roux}},\ }\emph {Distribution of
  the Ratio of Consecutive Level Spacings in Random Matrix Ensembles},\ \href
  {\doibase 10.1103/PhysRevLett.110.084101} {\bibfield  {journal} {\bibinfo
  {journal} {Phys. Rev. Lett.}\ }\textbf {\bibinfo {volume} {110}},\ \bibinfo
  {pages} {084101} (\bibinfo {year} {2013})}\BibitemShut {NoStop}%
\bibitem [{\citenamefont {Giraud}\ \emph {et~al.}(2022)\citenamefont {Giraud},
  \citenamefont {Mac\'e}, \citenamefont {Vernier},\ and\ \citenamefont
  {Alet}}]{GiraudAlet2020}%
  \BibitemOpen
  \bibfield  {author} {\bibinfo {author} {\bibfnamefont {O.}~\bibnamefont
  {Giraud}}, \bibinfo {author} {\bibfnamefont {N.}~\bibnamefont {Mac\'e}},
  \bibinfo {author} {\bibfnamefont {E.}~\bibnamefont {Vernier}}, \ and\
  \bibinfo {author} {\bibfnamefont {F.}~\bibnamefont {Alet}},\ }\emph {Probing
  Symmetries of Quantum Many-Body Systems through Gap Ratio Statistics},\ \href
  {\doibase 10.1103/PhysRevX.12.011006} {\bibfield  {journal} {\bibinfo
  {journal} {Phys. Rev. X}\ }\textbf {\bibinfo {volume} {12}},\ \bibinfo
  {pages} {011006} (\bibinfo {year} {2022})}\BibitemShut {NoStop}%
\bibitem [{\citenamefont {Rosenzweig}\ and\ \citenamefont
  {Porter}(1960)}]{RosenzweigPorter1960}%
  \BibitemOpen
  \bibfield  {author} {\bibinfo {author} {\bibfnamefont {N.}~\bibnamefont
  {Rosenzweig}}\ and\ \bibinfo {author} {\bibfnamefont {C.~E.}\ \bibnamefont
  {Porter}},\ }\emph {"Repulsion of Energy Levels" in Complex Atomic Spectra},\
  \href {\doibase 10.1103/PhysRev.120.1698} {\bibfield  {journal} {\bibinfo
  {journal} {Phys. Rev.}\ }\textbf {\bibinfo {volume} {120}},\ \bibinfo {pages}
  {1698} (\bibinfo {year} {1960})}\BibitemShut {NoStop}%
\bibitem [{\citenamefont {Berry}\ and\ \citenamefont
  {Robnik}(1984)}]{BerryRobnik1984}%
  \BibitemOpen
  \bibfield  {author} {\bibinfo {author} {\bibfnamefont {M.~V.}\ \bibnamefont
  {Berry}}\ and\ \bibinfo {author} {\bibfnamefont {M.}~\bibnamefont {Robnik}},\
  }\emph {Semiclassical level spacings when regular and chaotic orbits
  coexist},\ \href {\doibase 10.1088/0305-4470/17/12/013} {\bibfield  {journal}
  {\bibinfo  {journal} {Journal of Physics A: Mathematical and General}\
  }\textbf {\bibinfo {volume} {17}},\ \bibinfo {pages} {2413} (\bibinfo {year}
  {1984})}\BibitemShut {NoStop}%
\bibitem [{\citenamefont {Sun}\ \emph {et~al.}(2020)\citenamefont {Sun},
  \citenamefont {Yi-Xiang}, \citenamefont {Ye},\ and\ \citenamefont
  {Liu}}]{SunLiu2020}%
  \BibitemOpen
  \bibfield  {author} {\bibinfo {author} {\bibfnamefont {F.}~\bibnamefont
  {Sun}}, \bibinfo {author} {\bibfnamefont {Y.}~\bibnamefont {Yi-Xiang}},
  \bibinfo {author} {\bibfnamefont {J.}~\bibnamefont {Ye}}, \ and\ \bibinfo
  {author} {\bibfnamefont {W.-M.}\ \bibnamefont {Liu}},\ }\emph {Classification
  of the quantum chaos in colored Sachdev-Ye-Kitaev models},\ \href {\doibase
  10.1103/PhysRevD.101.026009} {\bibfield  {journal} {\bibinfo  {journal}
  {Phys. Rev. D}\ }\textbf {\bibinfo {volume} {101}},\ \bibinfo {pages}
  {026009} (\bibinfo {year} {2020})}\BibitemShut {NoStop}%
\end{thebibliography}
\end{document}